\documentclass[iop]{emulateapj}

\shorttitle{
The stellar halo in Gaia DR2}
\shortauthors{Haywood et al.}

\begin{document}

\title{In disguise or out of reach: 
first clues about \textit{in situ} and accreted stars in the stellar halo of the Milky Way  from Gaia~DR2
}

\author{M.~Haywood and P.~Di Matteo}

\affil{GEPI, Observatoire de Paris, PSL Research University, CNRS,  Sorbonne Paris Cit\'e, 5 place Jules Janssen, 92190 Meudon, France}
\email{misha.haywood@obspm.fr}

\author{M.~D.~Lehnert}
\affil{Sorbonne Universit\'e, CNRS UMR7095, Institut d’Astrophysique de Paris, 98 bis bd Arago, 75014 Paris, France}

\author{O.~Snaith}
\affil{School of Physics, Korea Institute for Advanced Study, 85 Hoegiro, 
Dongdaemun-gu, Seoul 02455, Republic of Korea}

\author{S.~Khoperskov
and
A.~G\'omez\altaffilmark}
\affil{GEPI, Observatoire de Paris, PSL Research University, CNRS, Sorbonne Paris Cit\'e, 5 place Jules Janssen, 92190 Meudon, France}

\begin{abstract}
We investigate the nature of the double color-magnitude sequence observed in the Gaia DR2 HR diagram of stars with high transverse velocities. The stars in the reddest-color sequence are likely dominated by the dynamically-hot tail of the thick disk population. Information from \citet{nissen2010} and from the APOGEE survey suggests that stars in the blue-color sequence have elemental abundance patterns that can be explained by this population having a relatively low star-formation efficiency during its formation. In dynamical and orbital spaces, such as the `Toomre diagram', the two sequences show a significant overlap, but with a tendency for stars on the blue-color sequence to dominate regions with no or retrograde rotation and high total orbital energy. In the plane defined by the maximal vertical excursion of the orbits versus their apocenters, stars of both sequences redistribute into discrete wedges. We conclude that stars which are typically assigned to the halo in the solar vicinity are actually both accreted stars lying along the blue sequence in the HR diagram, and the low rotational velocity tail of the old Galactic disk, possibly dynamically heated by past accretion events. Our results imply that a halo population formed \textit{in situ} and responsible for the early chemical enrichment prior to the formation of the thick disk is yet to be robustly identified, 
and that what has been defined as the stars of the \textit{in situ} stellar halo of 
the Galaxy may be in fact fossil records of its last significant merger.
\end{abstract}


\section{Introduction}

Over the past three decades, a consensus has developed that the
stellar halo of the Milky Way is composed of two populations of stars
-- those that were born in other galaxies and accreted and those that
were born \textit{in situ} during the early evolution of the Milky Way \citep[e.g.][]{searle1978,sommer1990,carollo2007}.
Although these two components have remained challenging to characterize,
we now have a widely-accepted picture whereby the \textit{in situ}
halo population is older, more metal-rich, dominates the stellar
density within $\sim$15kpc of the Galactic center, and has a slightly
enhanced mean rotation rate compared to the accreted halo population
\citep[e.g.,][]{carollo2007}. While a considerable amount of effort is
currently being expended investigating stellar streams that are
expected to fill the outer stellar halo, the inner halo population is
still lacking a proper characterization, and its evolutionary connection
with the thick disk is essentially unknown.

The second data release \citep{GaiaBrown2018} of the European Space
Agency's Gaia mission \citep{GaiaPrusti2016} provides superb astrometric
parameters, radial velocities and photometry for a large number of stars.
First results have already detected a number of streams and kinematic
groups \citep{malhan2018, koppelman2018}.  Inspection of the Gaia HR
diagram (hereafter HRD) of stars with high total or tangential velocities
has shown two parallel color sequences \citep{GaiaBabusiaux2018}, which
were attributed to stars in the thick disk and stellar halo. Here,
we concentrate on understanding the origin of these two sequences and
in particular the nature of the blue-color sequence.  The next section
describes our selection from the Gaia archive, Section~\ref{sec:ns}
analyzes the sample of stars from \cite{nissen2010} specifically with
respect to where stars in this sample lie in these two sequences and
thus bringing insights into the characteristics of the blue-color
sequence. In Section~\ref{sec:kin}, we present their main kinematic
and orbital properties and we discuss our results and summarize our main
conclusions in Section~\ref{sec:discussion}.

\begin{figure*}
\includegraphics[clip=true, trim = 4mm 0mm 14mm 3mm, width=6.cm]{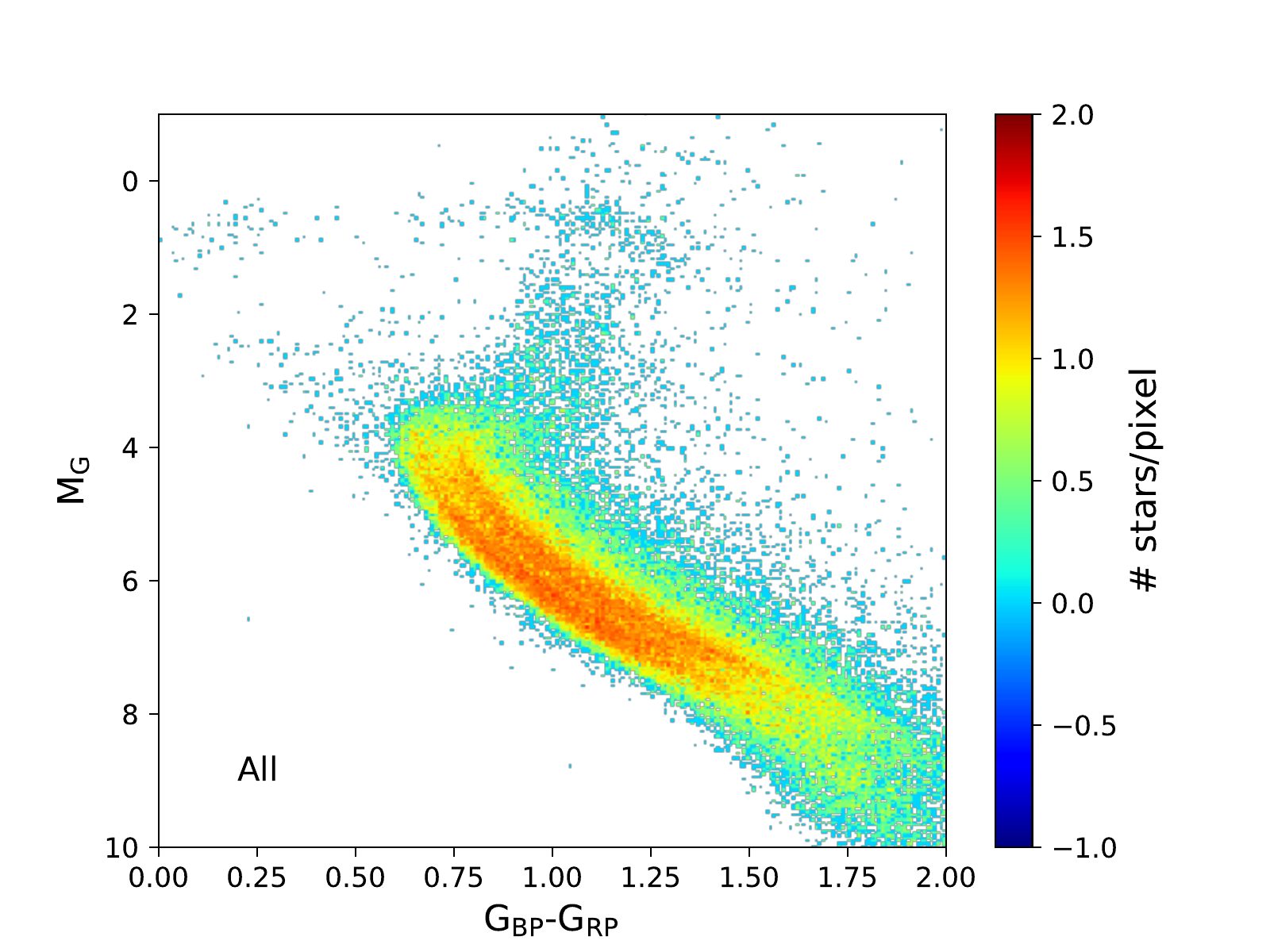}
\includegraphics[clip=true, trim = 4mm 0mm 14mm 3mm, width=6.cm]{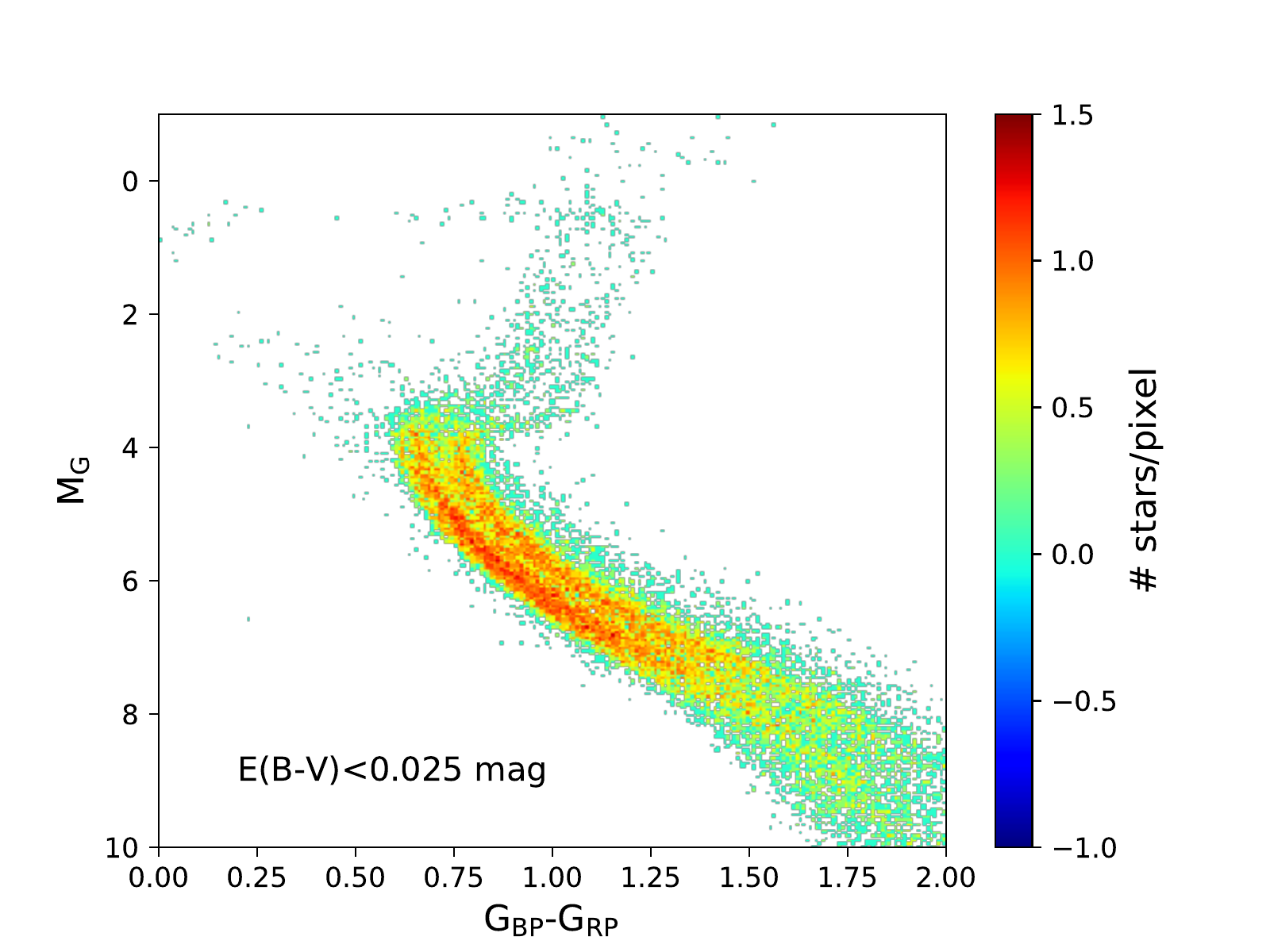}
\includegraphics[clip=true, trim = 4mm 0mm 14mm 3mm, width=6.cm]{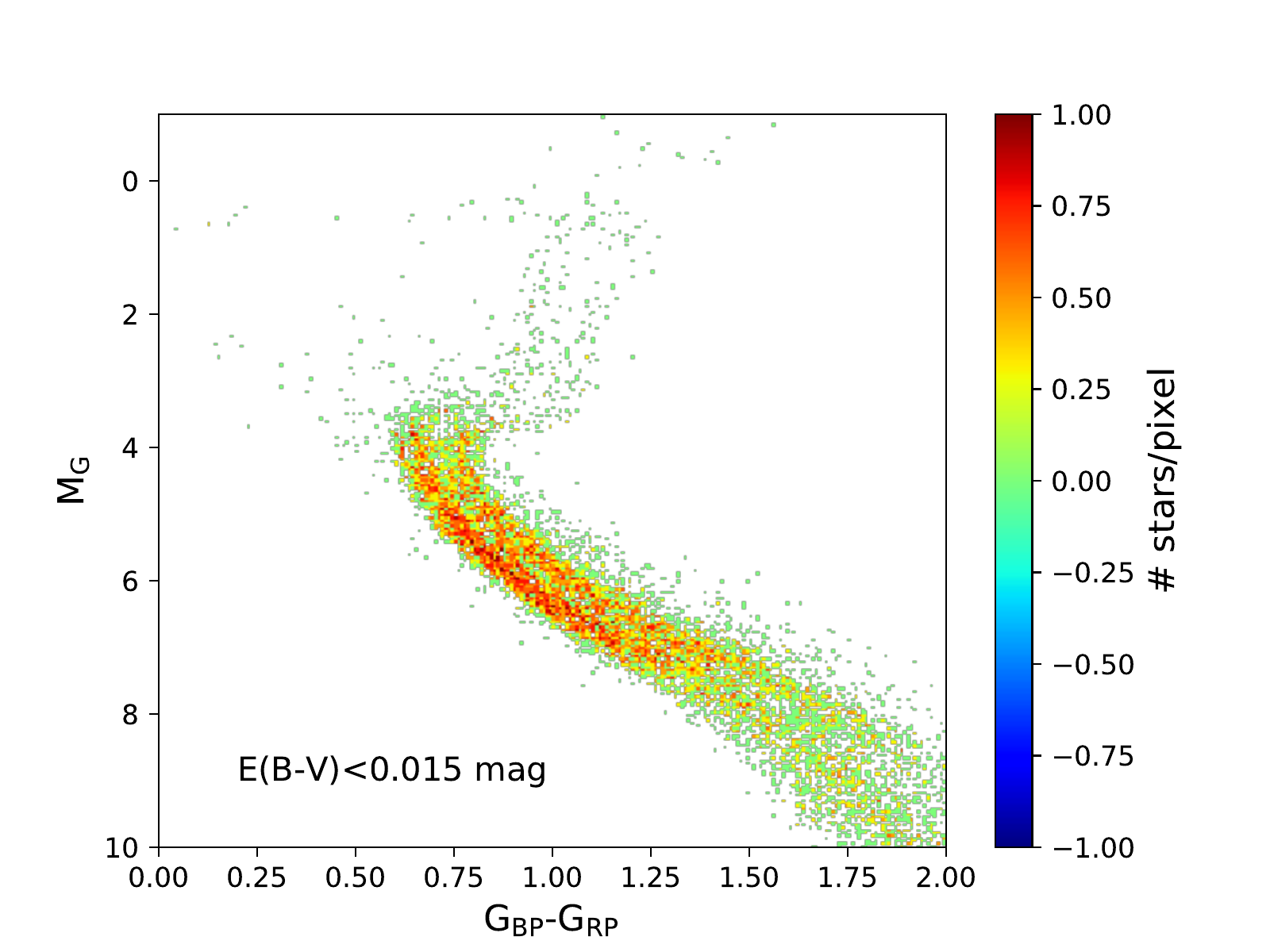}
\caption{HRD of stars selected in Section~\ref{sec:obs}, with different
tolerances in the reddening. The two, blue- and red-color sequences, are well
separated.  \label{fig1}}
\end{figure*}

\section{Clues from Nissen \& Schuster (2010)}\label{sec:ns}

\subsection{Data}\label{sec:obs}

We select stars in the Gaia archive having a tangential velocity
greater than 200km.$s\rm ^{-1}$, parallax errors $\sigma_{\pi}/\pi
< 0.1$, $\pi>1$~mas, G$<$17, and the filters described in
\cite{GaiaBabusiaux2018}. This brings us 77107 stars displayed in
the HRD of Fig.~\ref{fig1} (left  panel).  This figure shows that the
HRD  separates in two sequences, as in \cite{GaiaBabusiaux2018}, and
which we designate below as the blue and the red sequence (BS and RS,
respectively).

In order to clean the HRD, we select stars with various levels of
tolerance on the interstellar reddening. We adopt the interstellar
reddening estimates from the map of \cite{lallement2018}, which is well adapted for stars
nearer than 1~kpc.  Fig. \ref{fig1}
(middle and right panels) shows the HRD obtained by selecting stars that
have levels of reddening below 0.025 and 0.015~mag, providing respectively
28210 and 12620 objects.

\begin{figure}
\includegraphics[clip=true, trim = 4mm 0mm 4mm 3mm, width=8.cm]{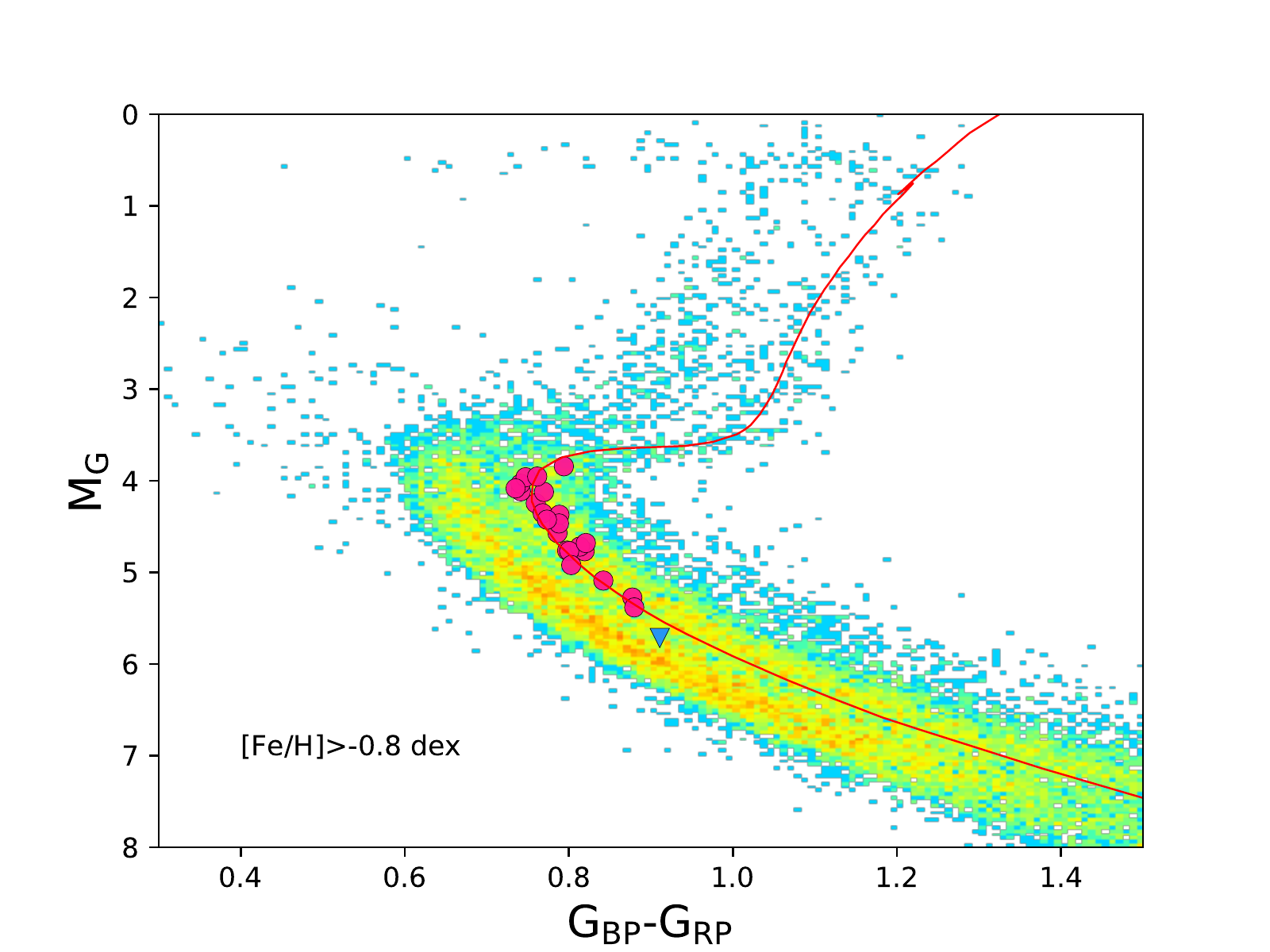}
\includegraphics[clip=true, trim = 4mm 0mm 4mm 3mm,width=8.cm]{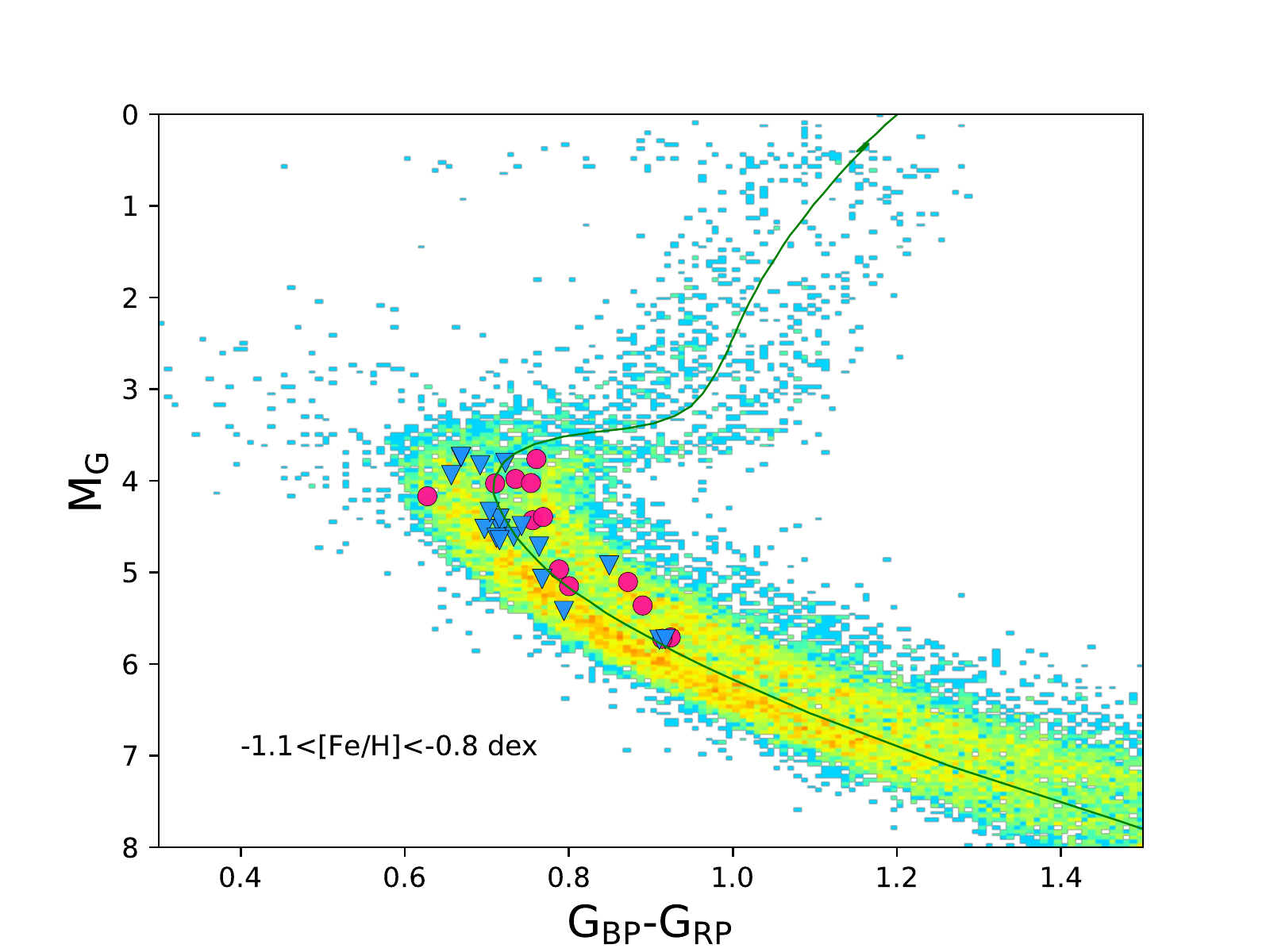}
\includegraphics[clip=true, trim = 4mm 0mm 4mm 3mm,width=8.cm]{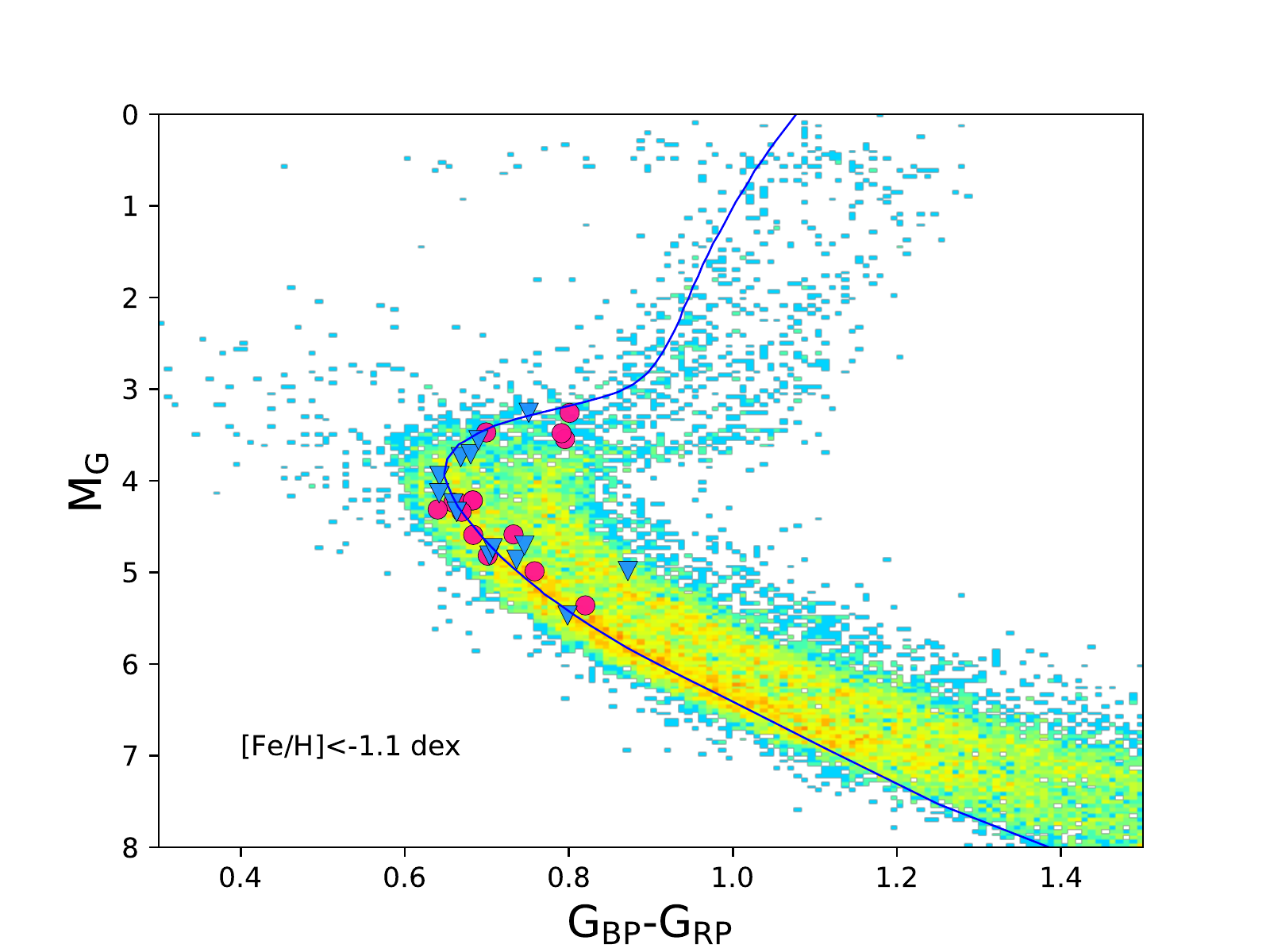}
\caption{Overplotted on the Gaia HRD, stars from Nissen \& Schuster (2010) with: [Fe/H]$>$-0.8 (\emph{top panel}), -1.1$<$[Fe/H]$<$-0.8 (\emph{middle panel}), [Fe/H]$<$-1.1 (\emph{bottom panel}). Thick disk and high-$\alpha$ halo stars from NS are represented with red circles, low-$\alpha$ halo stars from NS with blue triangles.
The isochrones from the PARSEC library have (top to bottom): (0.006 dex, 11 Gyr), (0.0048 dex, 11.5 Gyr), (0.0024 dex, 12 Gyr). 
}\label{fig:ns}
\end{figure}

\begin{figure}
\includegraphics[width=8.5cm]{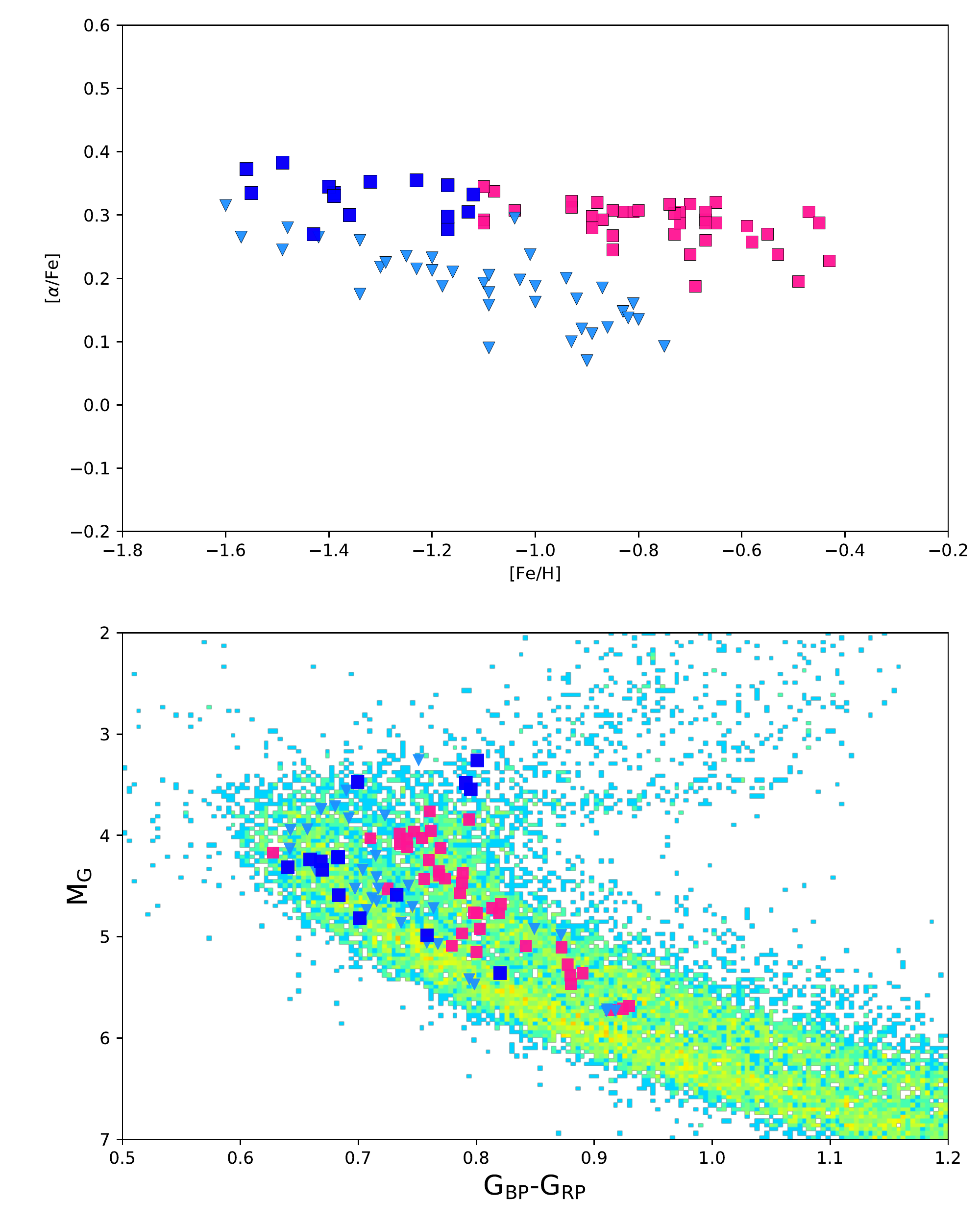}
	\caption{\textit{Top:} the low- and high alpha stars from \citet[][triangles and
squares]{nissen2010}. The high-$\alpha$ stars at [Fe/H]$<$-1.2 plotted{}
as blue squares may be assumed to be part of the same chemical sequence
\citep{hayes2018}. \textit{Bottom:} the resulting classification
is  more consistent with the belonging to the two sequences in the HRD,
see Fig.\ref{fig:ns}. 
}\label{fig:nsfehalpha}
\end{figure}

\subsection{Chemical abundances and HR diagram}

The sample of \cite{nissen2010} (hereafter NS) in three different metallicity
intervals is overplotted to the Gaia~HRD in Fig.~\ref{fig:ns}, together with 
isochrones from the PARSEC library \citep{marigo2017} at Z=(0.006, 0.0024, 0.0048) 
and ages(11, 11.5, 12.5) Gyr.  Note that the Gaia magnitudes were derived using the 
V,I magnitudes and color transformations provided by \cite{evans2018}.

This figure shows that the red and blue sequences are dominated by
metal-rich ([Fe/H]$>$-0.8) and metal-poor ([Fe/H]$<$-1.1) stars \citep[as already anticipated
in][]{GaiaBabusiaux2018}. Note that the spread in metallicity in each
sequence is significant, $\sim$0.6~dex. A cross-match between APOGEE \citep{majewski2017} and our Gaia sample yields 226 stars,
showing that there is a marked dip in the metallicity distribution of
stars at [Fe/H]$\sim$-1.0 \citep{GaiaBabusiaux2018}. This dip
is also noticeable in the data of NS (their Fig.~1). The middle plot
of Fig.~\ref{fig:ns} shows that stars with -1.1$<$[Fe/H]$<$-0.8,
which brackets the dip in metallicity, fall between the two sequences
of the HRD. This implies that the clear separation between the two
sequences is the conspicuous consequence of the dip in the metallicity
distribution function (MDF). Moreover, the middle plot shows that,
at similar metallicities,  low-$\alpha$ stars tend to be bluer than
high-$\alpha$ ones, suggesting that they must be slightly younger, supporting
similar results from \citet{schuster2012}. On the contrary, the
bottom panel of Fig.~\ref{fig:ns} shows that low and high-$\alpha$
stars with [Fe/H]$<$-1.1 from the NS sample can both be found
along the BS stars. This is surprising, because -- given the tightness
of the blue-color sequence -- if the low and high-$\alpha$ stars at [Fe/H]$<$-1.1 were belonging to two different chemical evolution
sequences -- as they do at -1.1$<$[Fe/H]$<$-0.8 -- we would expect a difference in age that would reflect in the HRD,
as is seen for their more metal-rich counterparts (Fig.\ref{fig:ns}, middle plot).
Is it thus possible that all the stars with [Fe/H]$<$-1.1 -- both high or low $\alpha$ stars, because the distinction 
becomes less obvious below this metallicity -- are the same population of stars, and are not 
causally related to the thick disk? Could they instead be causally related to the
low-$\alpha$ sequence at higher metallicities, which is clearly distinct from the thick disk?

By analyzing the [Fe/H]-[Mg/Fe] distribution of APOGEE stars,
\citet{hayes2018} have shown that the separation between the
high and low-$\alpha$ stars is not horizontal -- it is not a
separation in $\alpha$-abundances -- but that high-$\alpha$ stars at
[Fe/H]$\lesssim$-1.1 and low-$\alpha$ stars at higher metallicities
form a unique chemical sequence. The findings of
\citet{hayes2018} strongly support our suggestion. Fig \ref{fig:nsfehalpha}\footnote{In the first version of this
article, G 81-02 was found to fall blueward to the blue sequence, due to a probably eroneous 
extinction estimate. P. Nissen informed us that his own estimate from $uvby$ photometry indicate
a reddening of E(G$_{\rm BP}$-G$_{\rm RP}$)=0.043 mag, 5 times less than ours. This seems to be much more consistent with 
the metallicity of this star (-0.69 dex), and has been used here.} 
shows the [Fe/H]-[$\alpha$/Fe] distribution
of the sample of \citet{nissen2010}, with the high-$\alpha$, low metallicity stars 
shown as blue square.  Indeed, these objects are
all within the BS. It is therefore reasonable to suggest
that the low-metallicity ([Fe/H]$<$-1.1) stars of NS and the low-$\alpha$
sample at higher metallicity are forming a unique abundance sequence. The gap in the MDF at
[Fe/H]$\sim$$-$1 could simply be a reflection
of the transition between two populations of stars: 
a population whose origin needs to be determined (see below) at [Fe/H]$<$-1, and the 
thick disk above this limit.  \citet{hayes2018} also mention that the transition between the two
sequences in the [Fe/H]-[$\alpha$/Fe] plane is more pronounced at this
metallicity \citep[see also][]{bonaca2017}.

\subsection{Kinematics and orbits}\label{sec:kin}

Of the sample of 28210 stars with E(B-V)$<$0.025~mag, 1973 stars have
full 3D velocity information in Gaia~DR2. 
In all the following,  we have assumed an in-plane distance of the Sun from the Galactic center $R_{\odot}=8.34$~kpc following \citet{reid14},  a height of the Sun above the Galactic plane  $z_{\odot}=27$~pc \citep{chen01}, a velocity for the Local Standard of Rest  $V\rm_{LSR}=240$~km.s$^{-1}$ \citep{reid14} and a peculiar velocity of the Sun with respect to the LSR,  $U_{\odot}=11.1$~km.s$^{-1}$, $V_{\odot}=12.24$~km.s$^{-1}$, $V_{\odot}=7.25$~km.s$^{-1}$, following \citet{schonrich10}.

\begin{figure} 
\includegraphics[clip=true, trim = 0mm 0mm 10mm 13mm,width=0.9\linewidth]{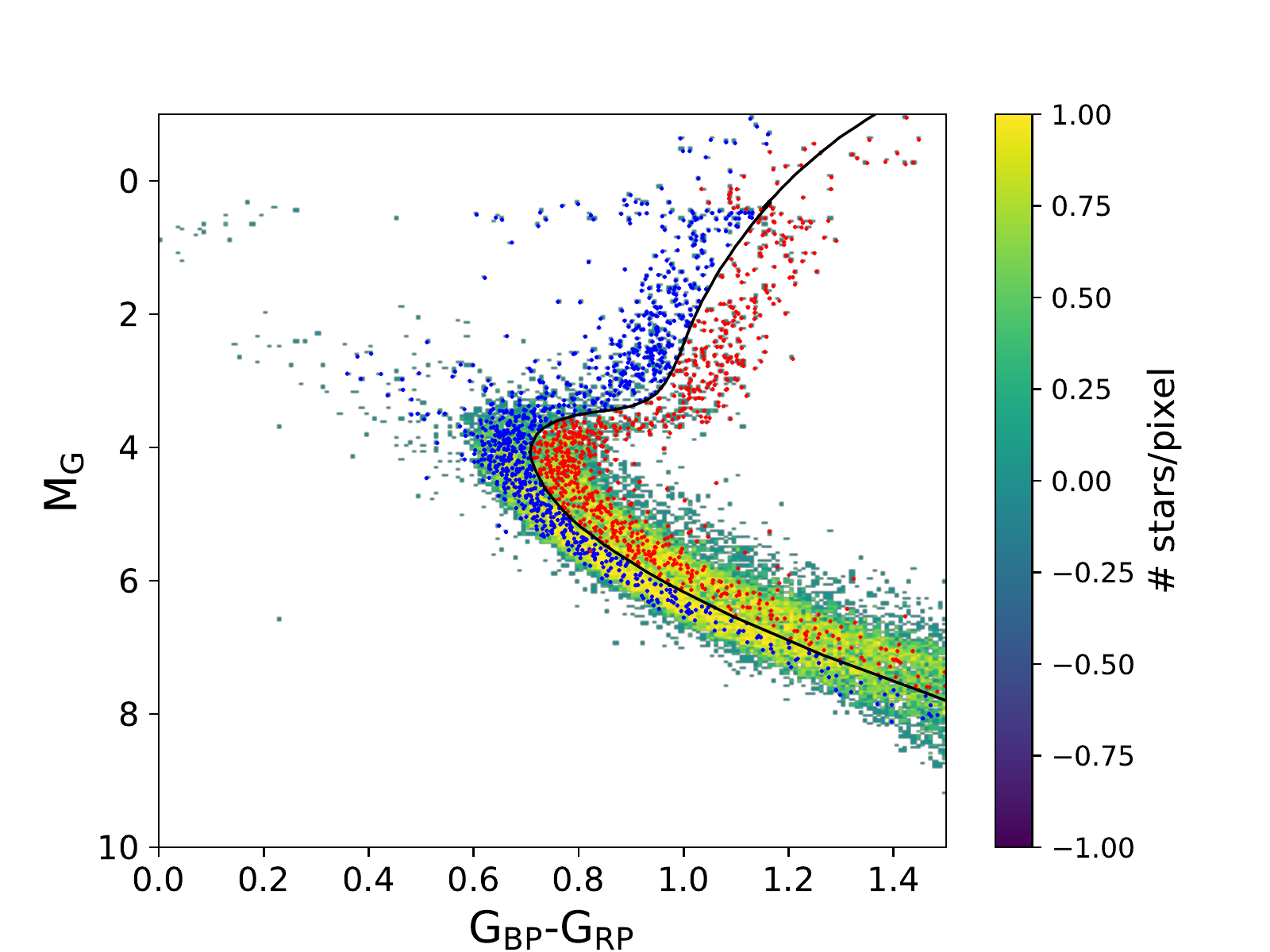}
\caption{Gaia HR diagram of the stars with 3D kinematics in our sample. We separate our sample 
in two groups (red and blue sequence) according to the position of each star relative to the 
isochrone. The isochrone is the same as the one in Fig. \ref{fig:ns}, middle plot.} 
\label{fig:sep}
\end{figure}

Figure \ref{fig:sep} show the HR diagram of the subsample of stars with 3D kinematics, and how we separate them using the 
isochrone.

\begin{figure} 
\includegraphics[clip=true, trim = 0mm 0mm 10mm 13mm,width=0.9\linewidth]{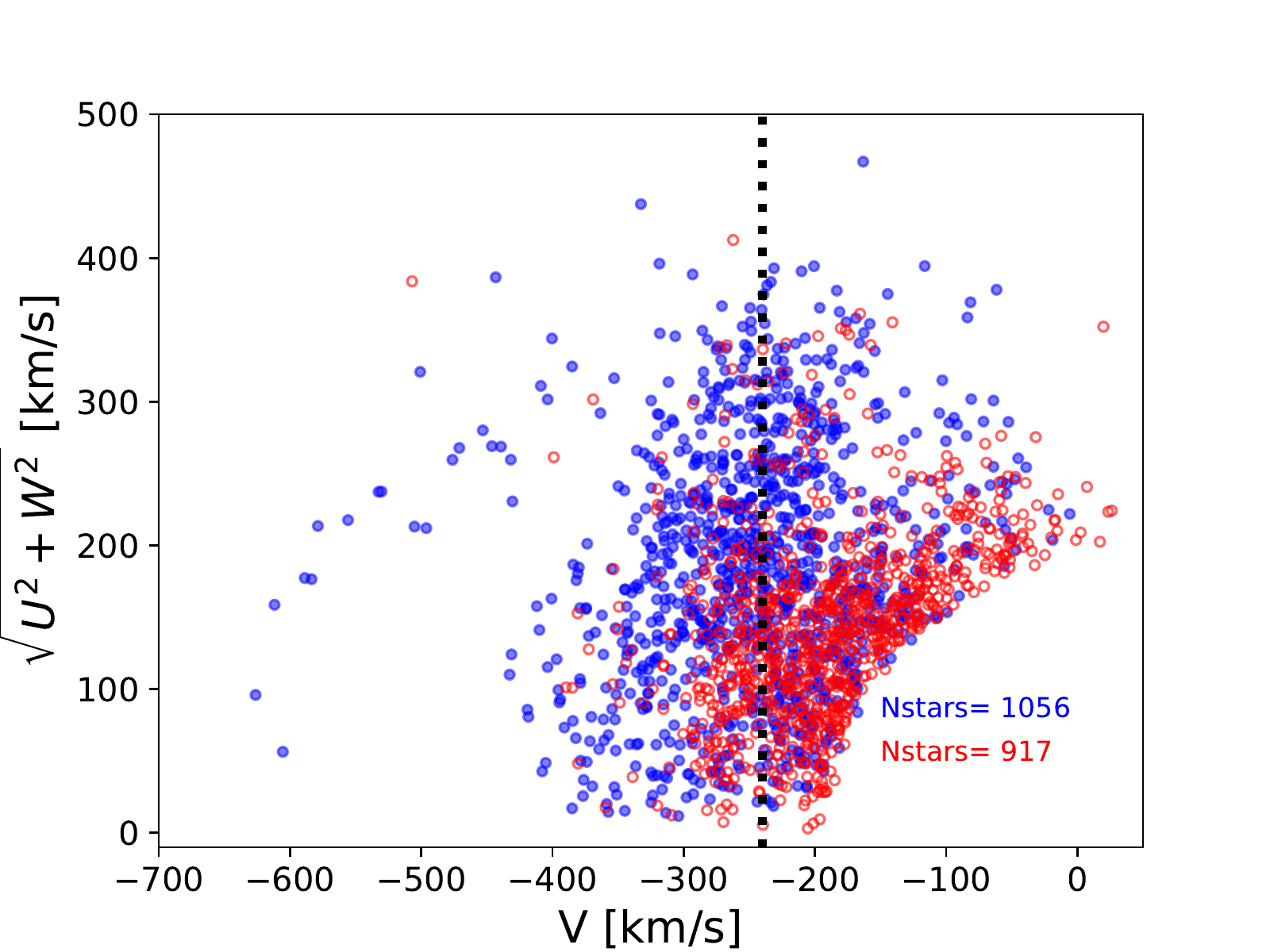} 
\includegraphics[clip=true,trim = 0mm 0mm 10mm 13mm, width=0.9\linewidth]{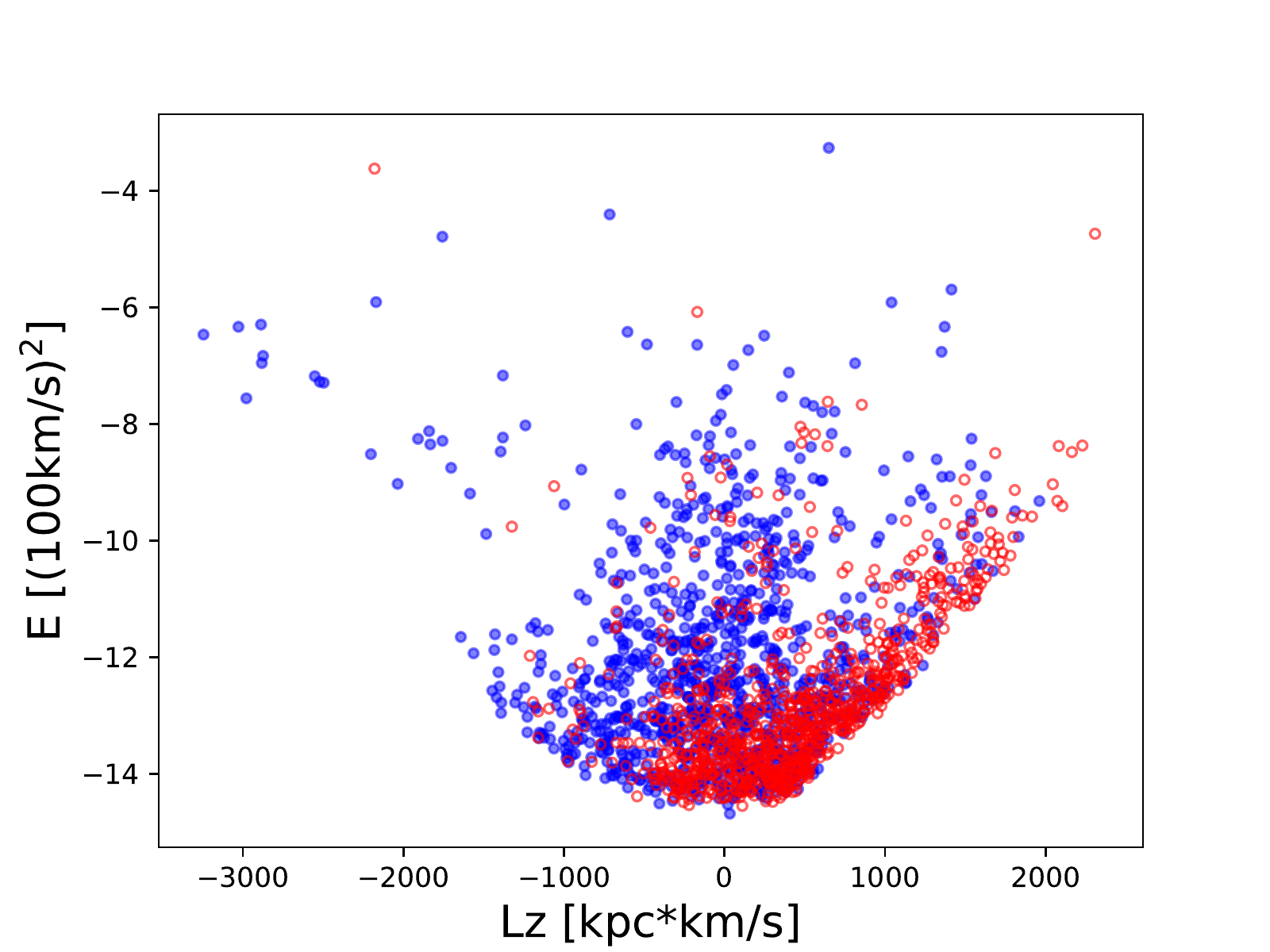}
\includegraphics[clip=true, trim = 0mm 0mm 10mm 13mm,width=0.9\linewidth]{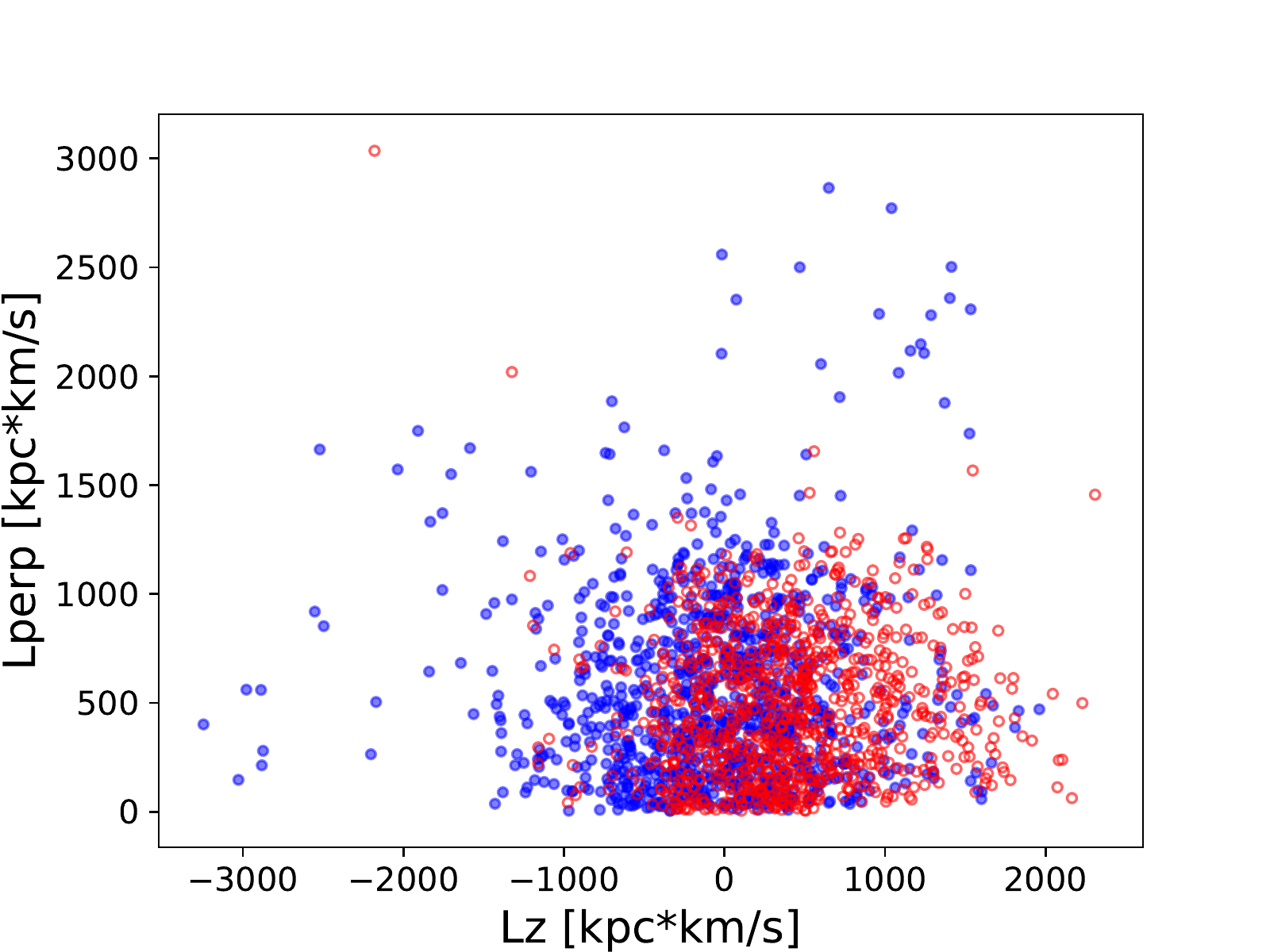} 
\includegraphics[clip=true, trim =0mm 0mm 10mm 13mm, width=0.9\linewidth]{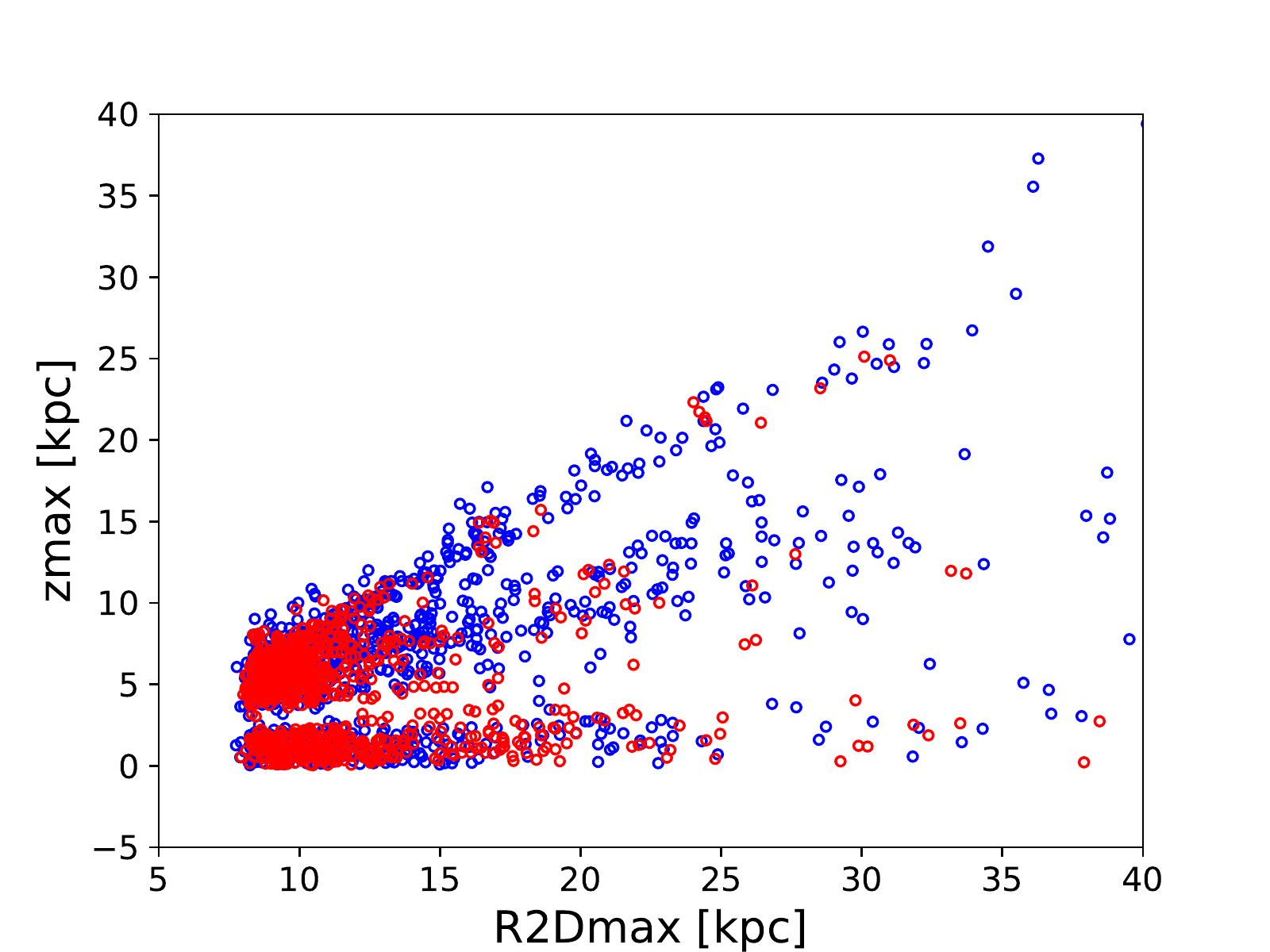}
\caption{\emph{From top to bottom:} Toomre diagram, $E-L_z$,
$L_z-L_\mathrm{perp}$ and $z_\mathrm{max}-R_\mathrm{2D,max}$ planes
for all stars in our sample with full 3D velocities in Gaia~DR2. The dashed line in the top plot indicates V=-V$\rm_{LSR}$.
In the bottom panel, only stars with $R_\mathrm{2D,max}$ $\lesssim$ 40~kpc have been plotted.
In all panels, stars on the blue sequence are shown as filled blue
dots, stars on the red sequence with empty red circles.} \label{kins}
\end{figure}

\begin{figure*} 
\includegraphics[clip=true, trim = 0mm 0mm 10mm 13mm,width=0.4\linewidth]{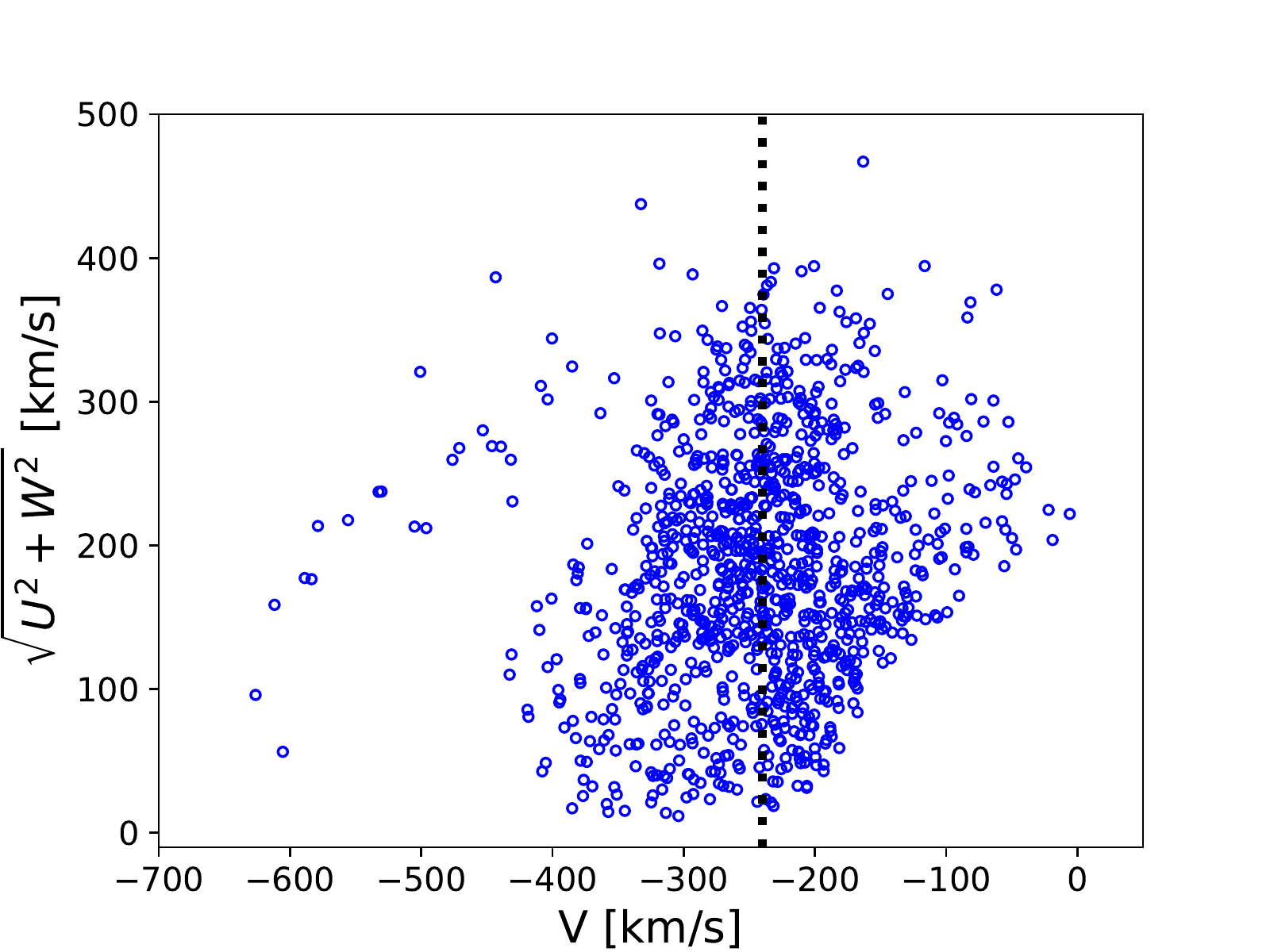}
\includegraphics[clip=true, trim = 0mm 0mm 10mm 13mm,width=0.4\linewidth]{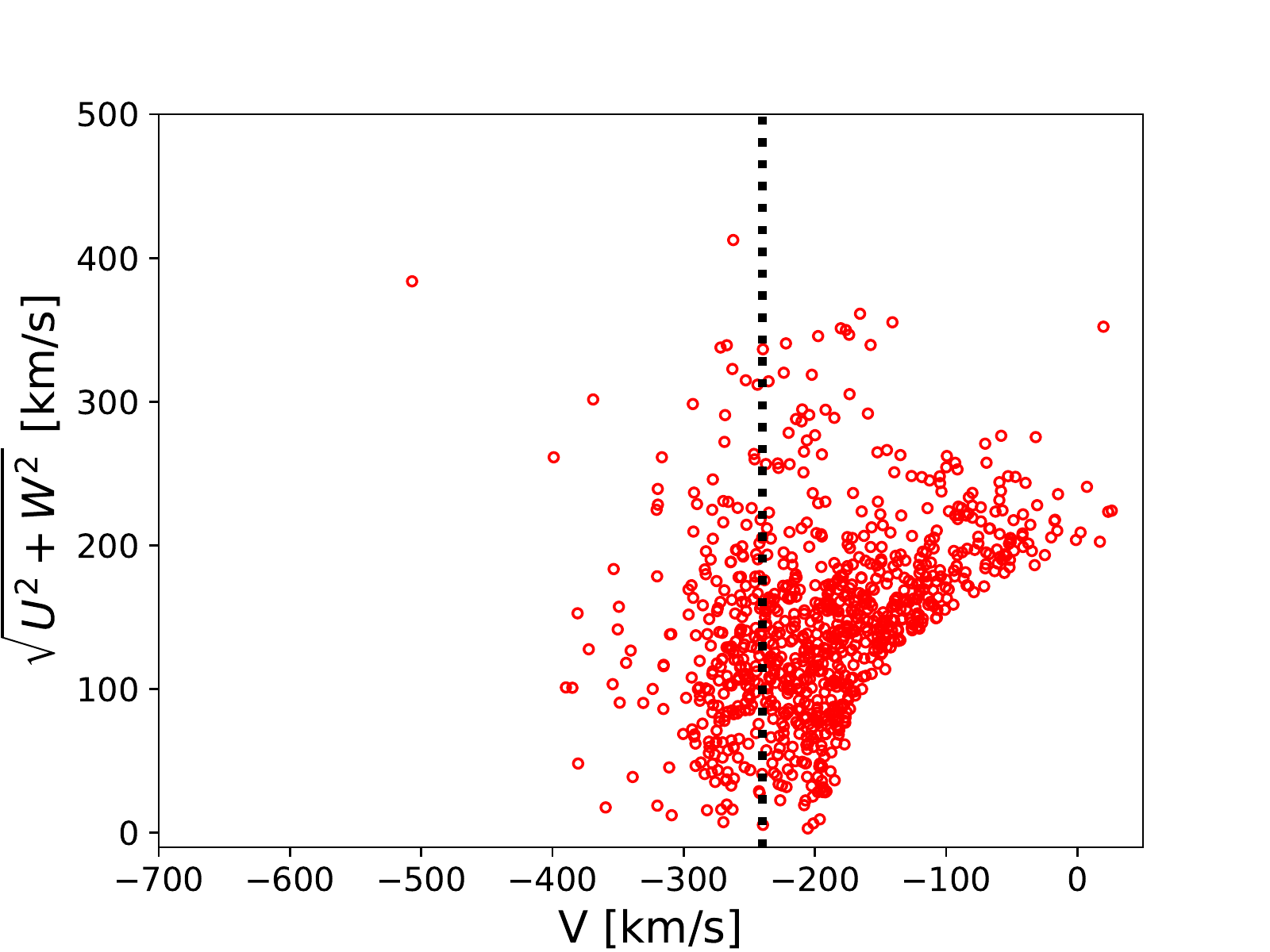}\\
\includegraphics[clip=true, trim = 0mm 0mm 10mm 13mm,width=0.4\linewidth]{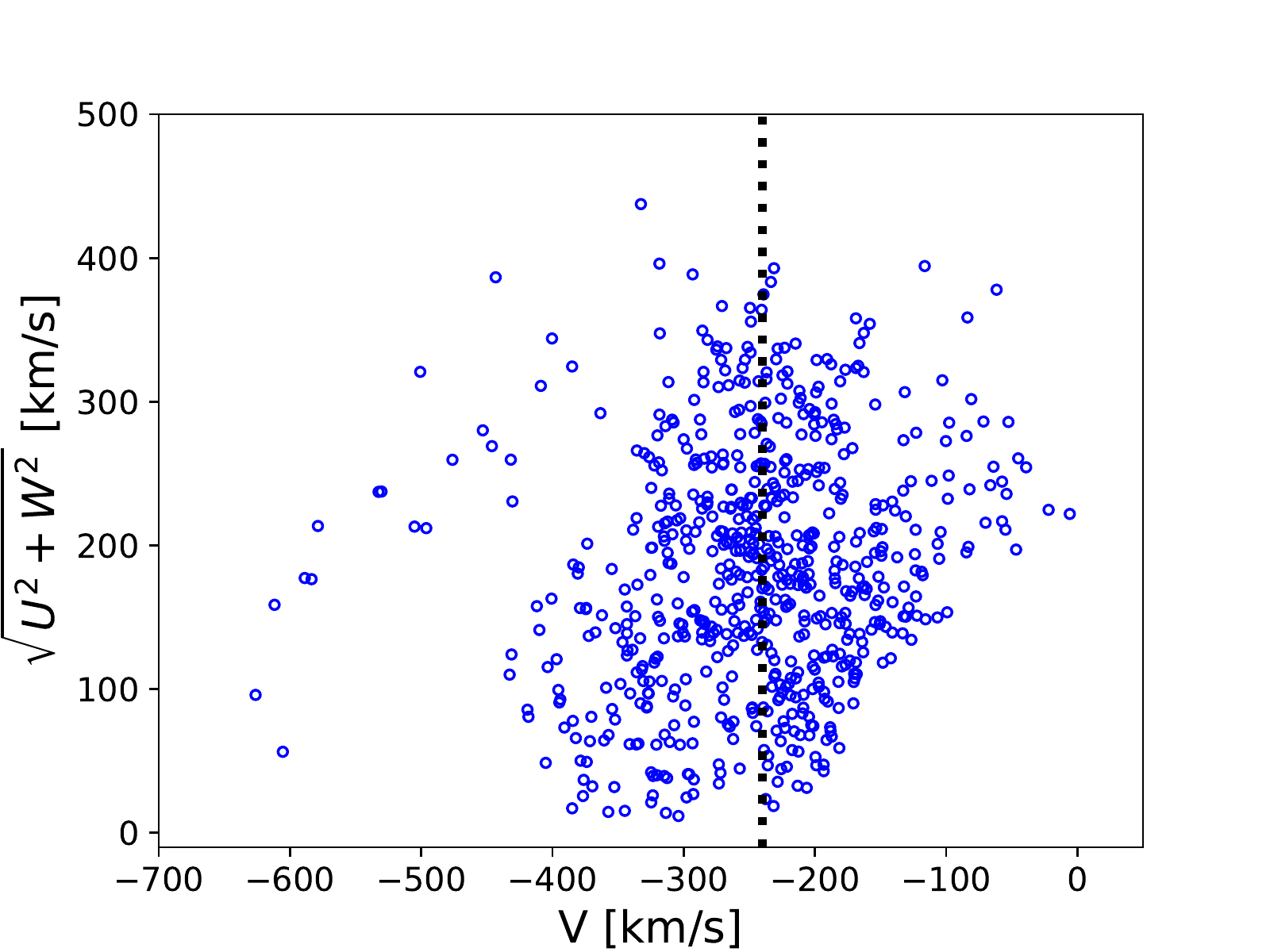} 
\includegraphics[clip=true, trim = 0mm 0mm 10mm 13mm,width=0.4\linewidth]{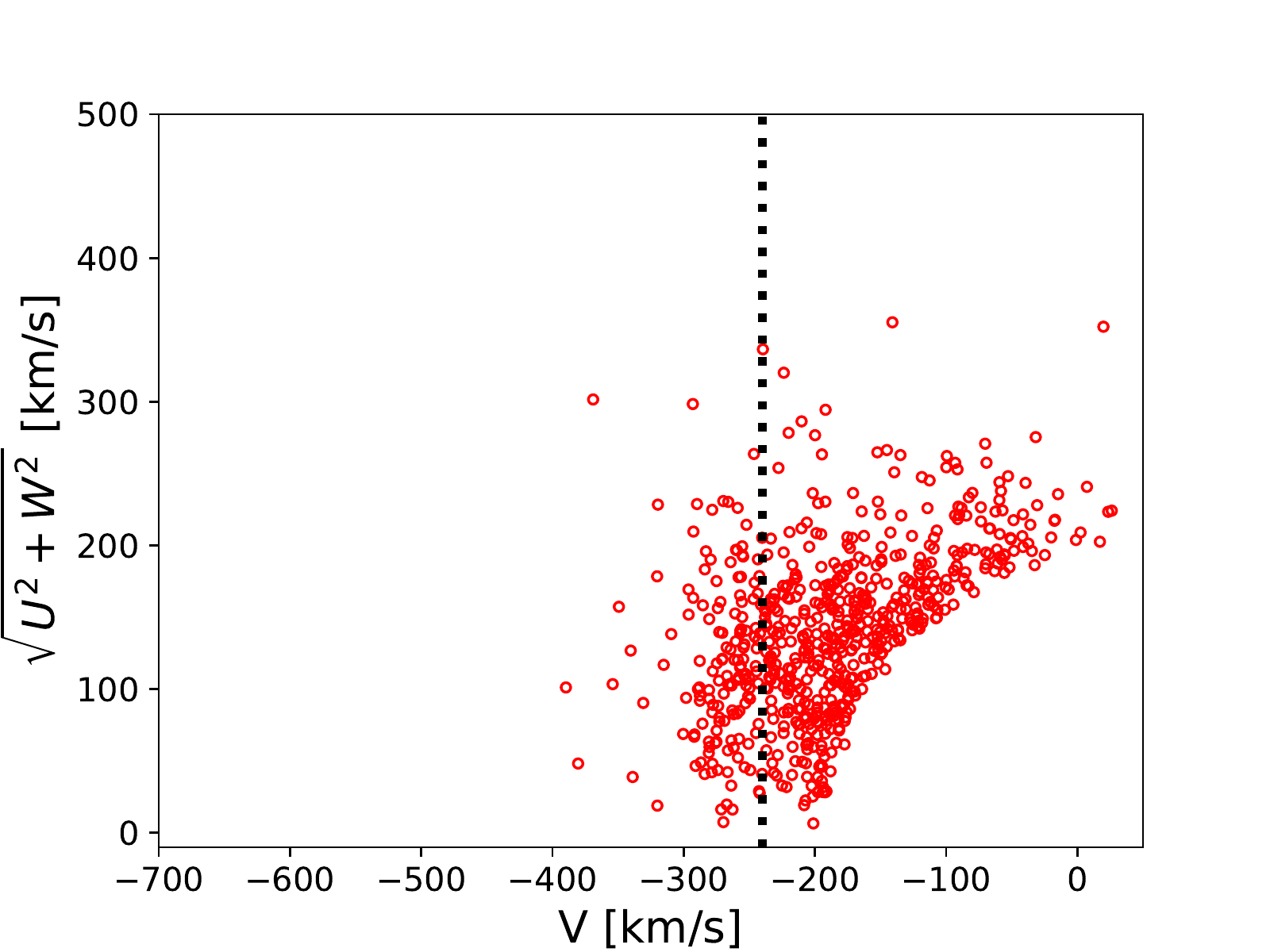} 
\caption{\emph{Top:} Toomre diagram shown separately for stars on the blue and red sequences, with the same 
separation as in Fig. \ref{kins}. \emph{Bottom:} the separation is now done by selecting stars bluer by -0.05 than the isochrone, and redder by + 0.05 mag for the stars in red, so the two samples are separated by 0.1 mag in color, decreasing
possible contamination between the two. See text for comment.
} \label{fig:testsep}
\end{figure*}

The distributions of our stars in kinematic spaces is given
in Fig.~\ref{kins}.  It shows two overlapping structures: a
vertical plume which extends to high energies and was already noted
in \cite[][their Fig. 5]{GaiaKatz2018}, and is composed of stars on the blue
sequence, as noted by \cite{koppelman2018}, and the high energy tail
of the disk, which extends to higher rotation, and contains stars on the
red sequence. The two sequences overlap in the Toomre diagram, however
the majority of stars on the RS, $\approx 75\%$, shows prograde rotation
and lies in the region of the diagram that is compatible with the slow
rotating tail of the thick disc (Fig.~\ref{kins}). The remaining fraction,
1/4 of red-color sequence stars, however, has retrograde orbits, lagging
the LSR by $\sim$400~km s$^{-1}$ or less. Except for 13 stars with very
significant retrograde motions, $V < -500$~km/s, that we will discuss
in the following, stars on the BS are distributed over the same area as
the RS stars in the Toomre diagram, but in different proportions. In
particular, they dominate the extended vertical plume at null or 
retrograde rotation, with about $65\%$ of BS stars lying with in the
region with $-350 \rm km/s \le \rm V \le -200 \rm km/s$,  while $26\%$
have $\rm V > -200 \rm km/s$ and the remaining fraction $\rm V < -350
\rm km/s$.  Remarkably, stars of the BS with prograde orbits clearly
overlap with those of RS stars.

In order to understand how these results are sensitive to the adopted
classification on the red and blue sequences, we studied how the
Toomre diagram changes when adopting a stricter separation. Stars
have been selected respectively +0.05 redder and -0.05 bluer than the
isochrone. The result is shown in Fig. \ref{fig:testsep}, and shows that
the effect is slightly different for the two populations: the contours of
the distribution of stars on the blue sequence remain basically the same, with even
the stretch of stars at strongly retrograde orbits being still present
in the selection.  On the contrary, several stars of the red sequence
on high energy orbits have been removed, suggesting that the vertical
plume, at least at the highest energies, could be almost exclusively
populated by stars on the blue sequence.  While a stricter
separation preferentially removes stars on high energy orbits from
the red sequence, we note that apart from this effect, the
main characteristics of the two distributions are essentially unchanged
using the two different selections.

The same main structures visible in the Toomre diagram are found
also in the $E-L_z$ plane (Fig.~\ref{kins}), where the energy $E$
is the total energy of a star, defined as the sum of its kinetic
and gravitational potential energies. We assume an \citet{allen1991} (hereafter AS)
Galactic mass model to estimate the latter. In this plane, we note
that the RS stars around $L_z = 0$ are predominately on very bound
orbits (i.e., low E).  In the $L_z-L_{perp}$ plane, where
$L_\mathrm{perp}=\sqrt{{L_x}^2+{L_y}^2}$, there is significant
overlap between the BS and RS within this space, except in two regions: (1)
the region where the most extreme counter-rotating stars lie, which
is composed of stars exclusively from the blue-color sequence; and
(2) an over-dense region roughly centered at
$(L_z,L_\mathrm{perp})$$\approx$$(1500, 2200)$~kpc$\cdot$km~s$^{-1}$,
which is very close to the region occupied by the Helmi stream
\citep[($L_z,L_\mathrm{perp}$)$\approx$ (1000, 2000)~kpc$\cdot$km~s$^{-1}$,
see][]{helmi99}, which is also exclusively made of stars on the BS.

Finally, for all stars with full 3D velocity information, we have
reconstructed their orbital parameters, by integrating their orbits
over the last 6~Gyr, using four different Galaxy potentials: the
axisymmetric potential of \citet{allen1991}, the two axisymmetric
potentials, including a thick disc component, in \citet{pouliasis2017}, and 
the MWpotential2014 from \cite{bovy2015}.
In Fig~\ref{kins}, we show their distribution in the
$z_\mathrm{max}-R_\mathrm{2D,max}$ plane, where $z_\mathrm{max}$
is the maximum height that stars reach above or below the Galactic
plane and, $R_\mathrm{2D,max}$, is the apocenter of their orbit
projected on the Galactic plane. For clarity, we show in this plot
only the orbital parameters that have been derived by integrating
the orbits in the \citet{allen1991} potential, and we discuss
below the robustness of this analysis when the other three potentials are used.  Predominately, the
stars on the RS have $R_\mathrm{2D,max}$ within 20~kpc of the
Galactic center, while the orbits of stars on the BS can extend
much further out than 20~kpc. A striking feature is that the stars
in our sample are not homogeneously distributed in this plane, but
define three distinct diagonal ``wedges'', with $z_\mathrm{max}$
increasing with $R_\mathrm{2D,max}$.

Remarkably, one of these three patterns defines stars confined in
a relatively thin and flattened distribution, with a lack of stars
between 2$\le$$z_\mathrm{max}$$\le$4~kpc.  This lack of stars with
$z_\mathrm{max}$ in this range makes this region distinct from the
rest of the sample. Ten of the thirteen stars with very significant
retrograde motions, $V<-500$~km~s$^{-1}$, are found in this thin
flattened disc.  Note that the presence of two groups of halo stars,
a first with a flattened distribution and a second with more
vertically extended orbits was already noted by \citet[][see their
Fig.~8]{schuster2012}.  

\begin{figure} 
\includegraphics[clip=true, trim = 0mm 0mm 0mm 0mm,width=1.\linewidth]{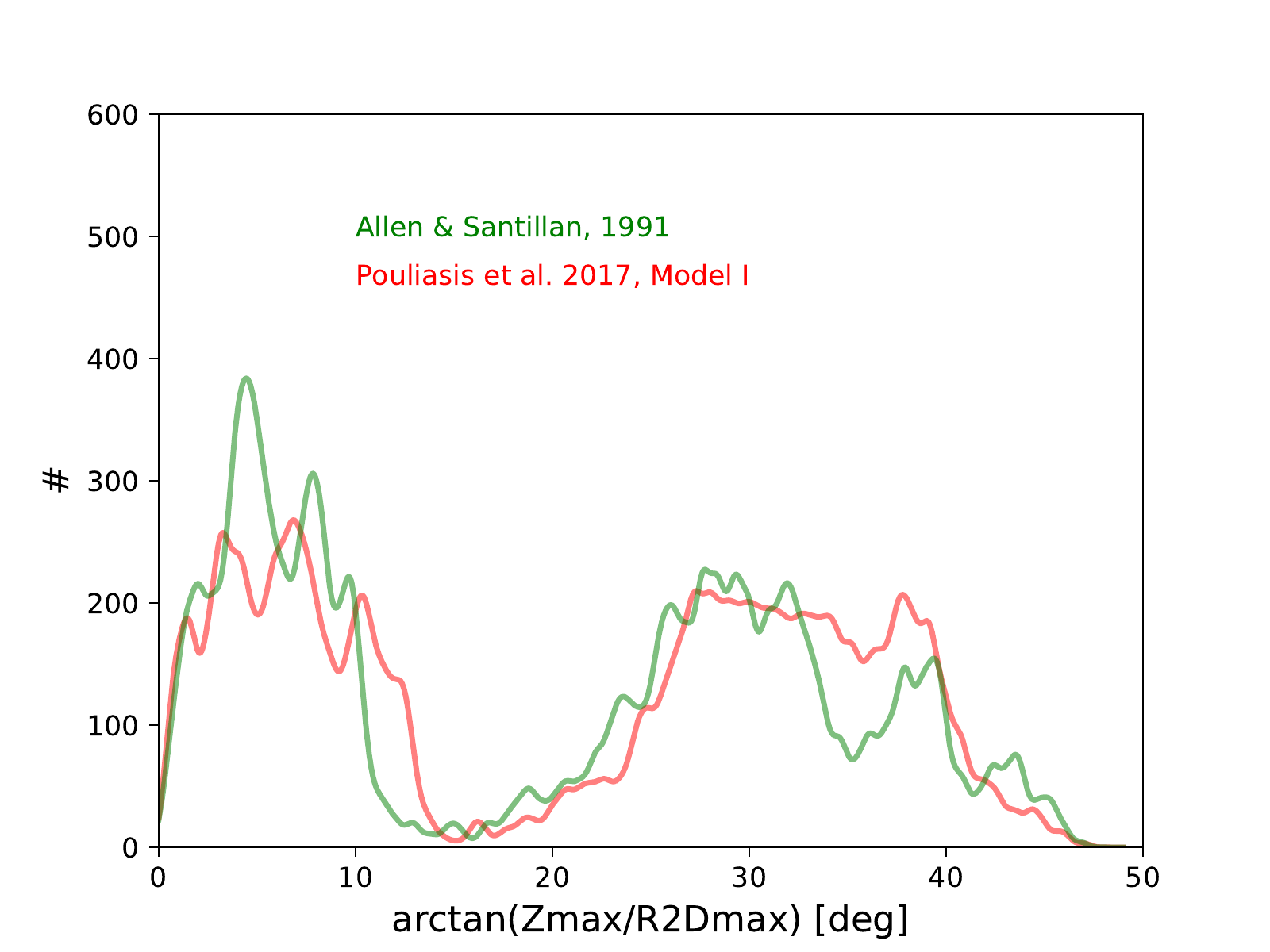} 
\includegraphics[clip=true, trim = 0mm 0mm 0mm 0mm,width=1.\linewidth]{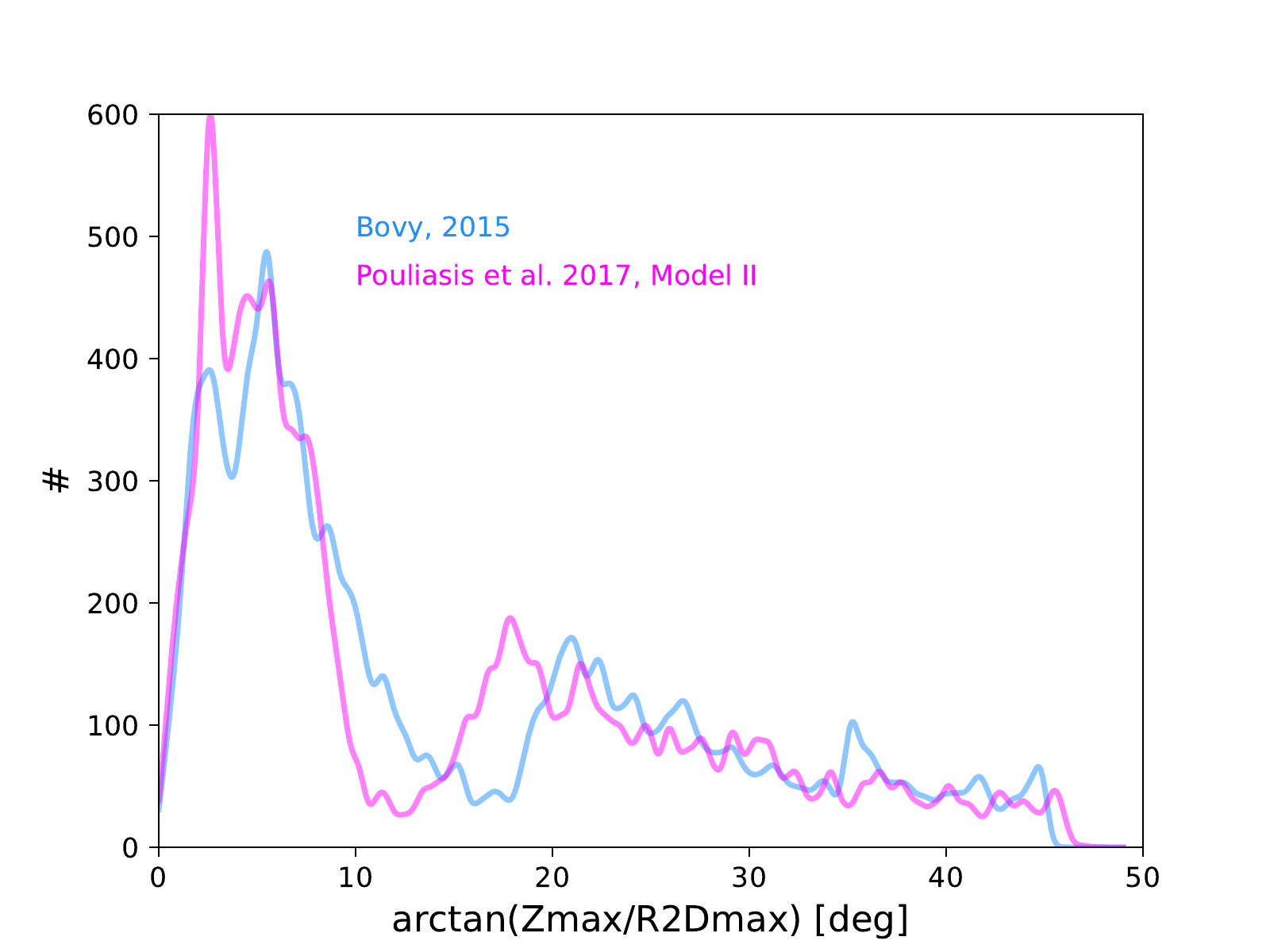} 
\caption{Distribution of angles (arctan(Zmax/R2Dmax)) obtained by calculating ten orbits for each star in four different 
Galactic potentials. The dip in the distribution (between 10 and 25$^\circ$) between the lower and intermediate wedges is 
found in all four potentials. The dip between the intermediate and upper wedges is not.} 
\label{fig:angles}
\end{figure}

In order to assess if the structures in the (RD2max, Zmax) plane
are robust to (1) the uncertainties in the parallax, proper motions and
radial velocity; and (2) the uncertainties in the mass distribution in the
Galaxy, we generated 10 random sets of these parameters for each star
around the observed values, using the 1-$\sigma$ uncertainties given
in the Gaia DR2 catalogue.  
The orbits were then integrated from the
U,V,W velocities in four different Galactic potentials described above.  
From the orbital parameters generated in all four potentials, we derived the angles 
defined by the arctan(Zmax/R2Dmax), and plot the resulting distributions in  
Figure~\ref{fig:angles}.

The potential of Model I in \cite{pouliasis2017} is similar to
the \cite{allen1991} potential, but includes a massive thick disk,
but since the overall mass distribution is fitted to the same rotation
curve, the overall potential is not very different, hence the distributions of 
Figure~\ref{fig:angles} (top) are very similar.
On the contrary, Model II in \cite{pouliasis2017} was fitted to a different rotation curve
\citep[see][for details]{pouliasis2017}, having a more massive dark matter halo,
and thus its distribution differs significantly from the previous two potentials, 
but is similar to the distribution obtained with the MWpotential2014 from \cite{bovy2015}.

A common feature seen in all
four distributions is the dip between 10-25$^\circ$ (Fig.~\ref{fig:angles}). It is wider in
the first two potentials, but is significant in all resulting distributions. The
gap between the intermediate and highest wedge is visible in only one
potential, but there seems to be nothing significant in the other three
potentials. We conclude that (only) the dip at 10-25$^\circ$ is significant.

\subsection{The effect of the kinematic selection}

\begin{figure} 
\includegraphics[clip=true, trim = 0mm 0mm 0mm 0mm,width=1.\linewidth]{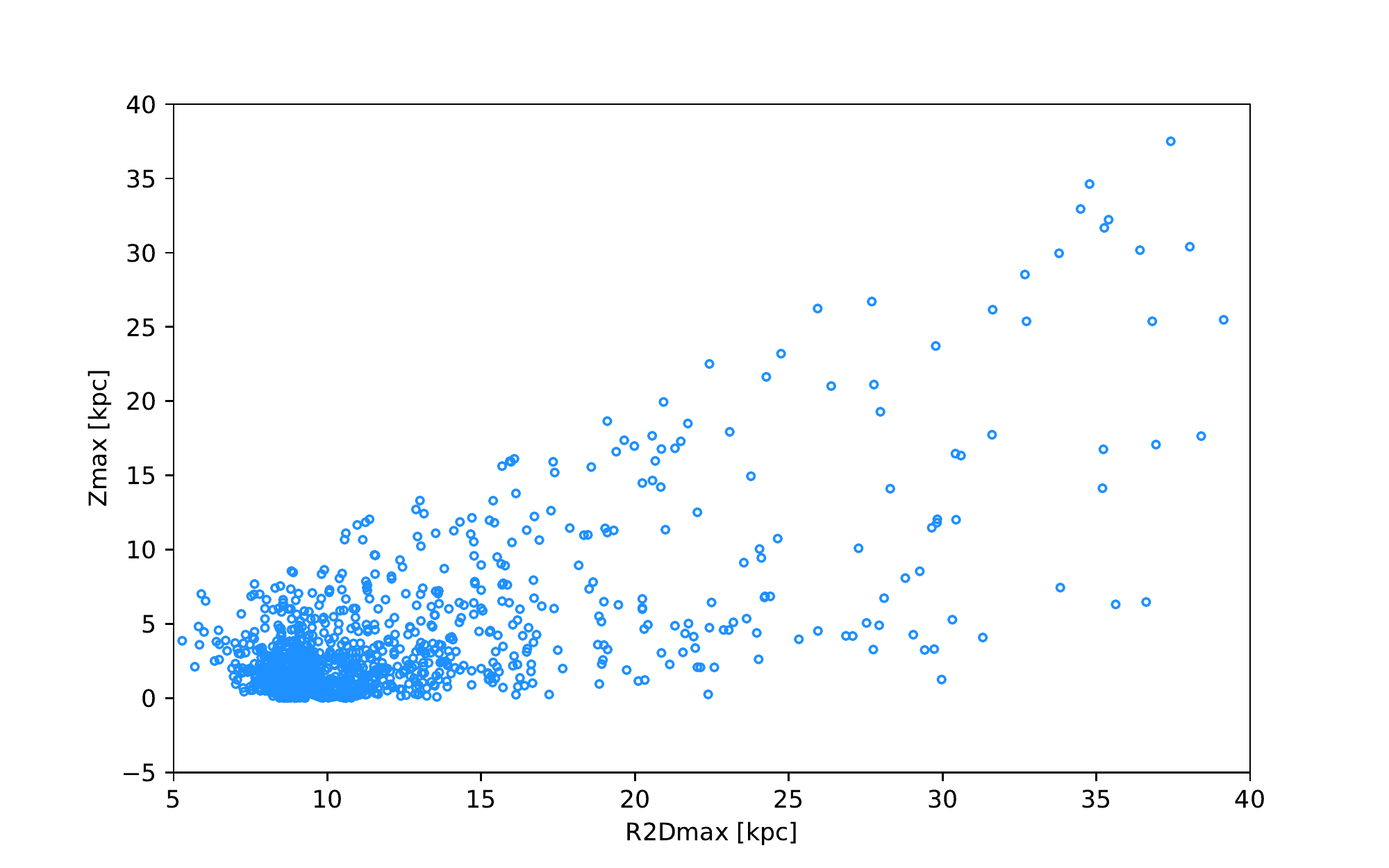} 
\includegraphics[clip=true, trim = 0mm 0mm 0mm 0mm,width=1.\linewidth]{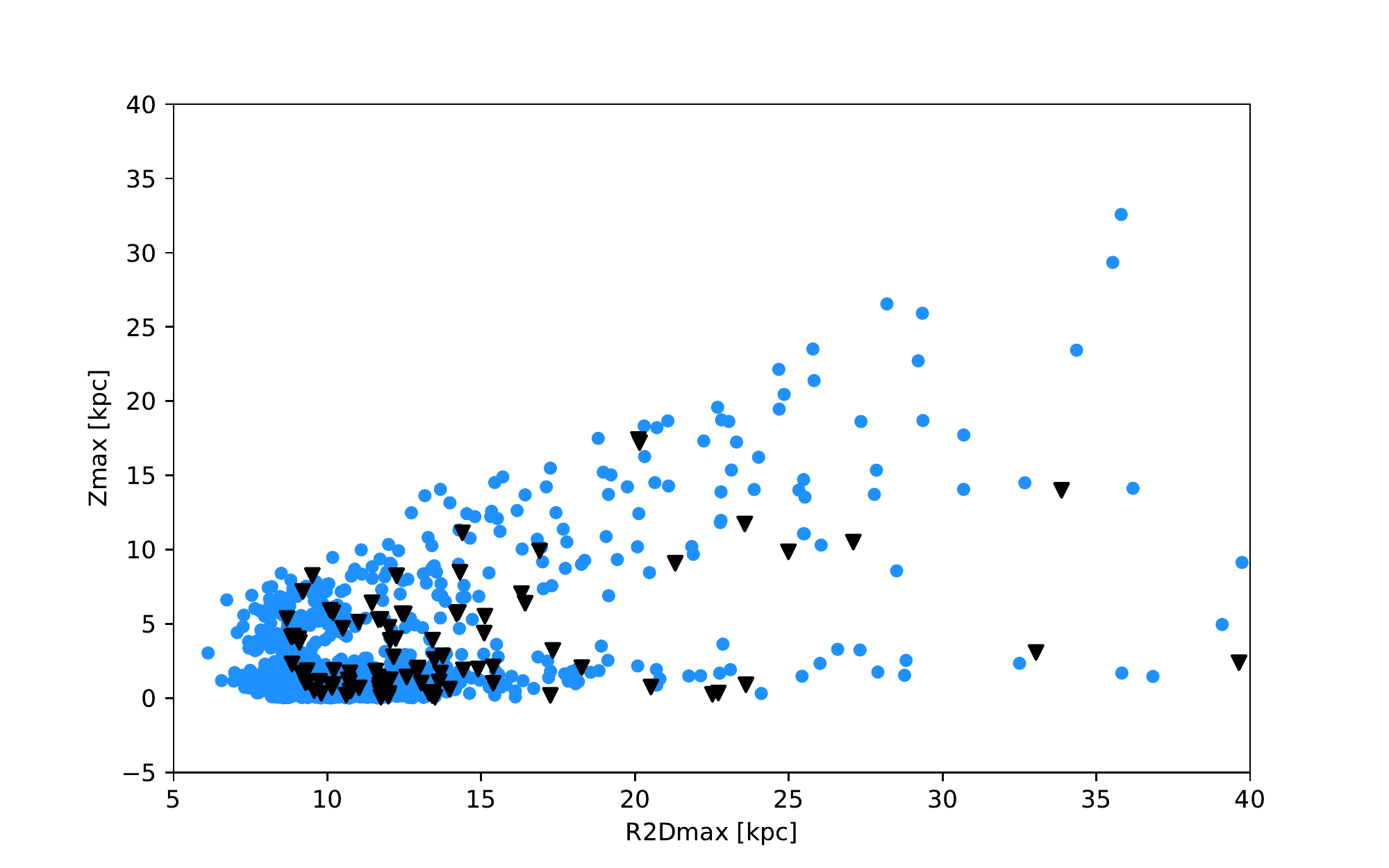} 
\caption{Top: The (R2Dmax,Zmax) distribution of stars from \cite{chiba2000} with the orbital parameters determined by the authors. The same sample with the orbital parameters recalculated with the AS potential and Gaia astrometric parameters and radial velocity, together with 
the sample of \cite{schuster2012} (black triangles), with their parameters derived from orbits calculated in the
barred potential.}\label{fig:chiba}
\end{figure}

In this study, we selected stars having tangential velocities higher than 200km.s$^{-1}$. 
Does the cut on tangential velocities affect our conclusions? One possible way to 
answer this question is to compare our results with samples that were not kinematically 
defined. To do so, we look at the dataset of \cite{chiba2000}, which 
has been designed to avoid kinematic bias. 
Fig.~\ref{fig:chiba} (top) shows the (R2Dmax,Zmax) plane adopting the orbital parameters published in 
\cite{chiba2000}, which shows that the structures
were already apparent in the (Zmax,R2Dmax) plane. 
Figure \ref{fig:chiba} (bottom) shows the same distribution
but with orbital parameters recalculated in the AS potential using
the Gaia DR2 astrometric parameters.  The separation between the lower and
intermediate wedges is clear.  The stars from \cite{schuster2012}, 
comprising halo stars with total velocities greater than 180km.$s^{-1}$ but also 
thick disk stars (see Fig. 3 in \cite{nissen2010}), 
are also overplotted (black triangles), with most objects populating the low and intermediate
wedges, with a clear separation between the two. Note that the orbital parameters of the \cite{schuster2012} stars
shown in Fig. \ref{fig:chiba} (bottom) correspond to those derived in a asymmetric barred potential
(see \cite{schuster2012} for details).

\section{The APOGEE-Gaia sample}\label{sec:apogee}

\begin{figure} 
\includegraphics[clip=true, trim = 20mm 0mm 10mm 13mm,width=1.1\linewidth]{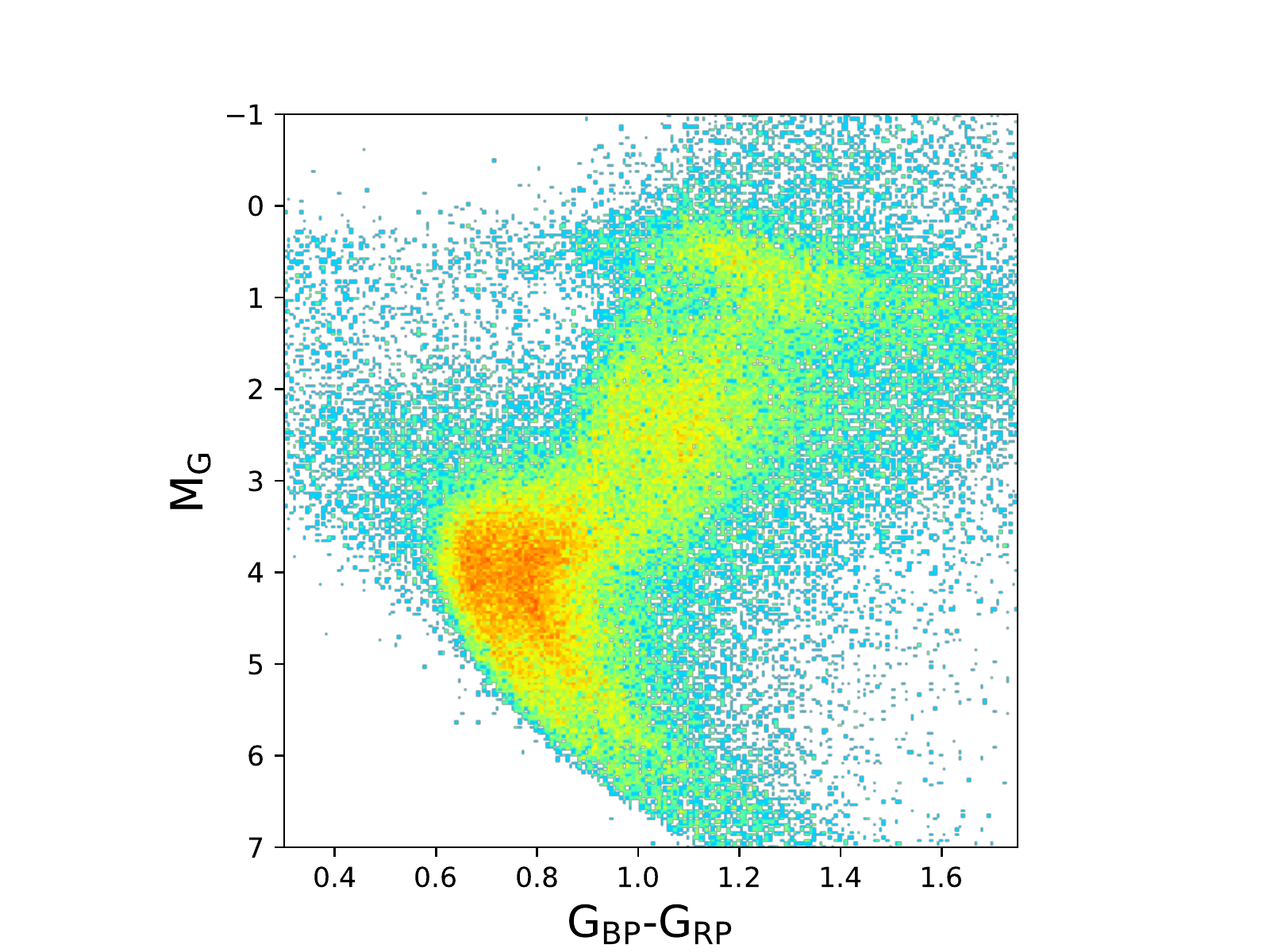}
\includegraphics[clip=true, trim = 20mm 0mm 10mm 13mm,width=1.1\linewidth]{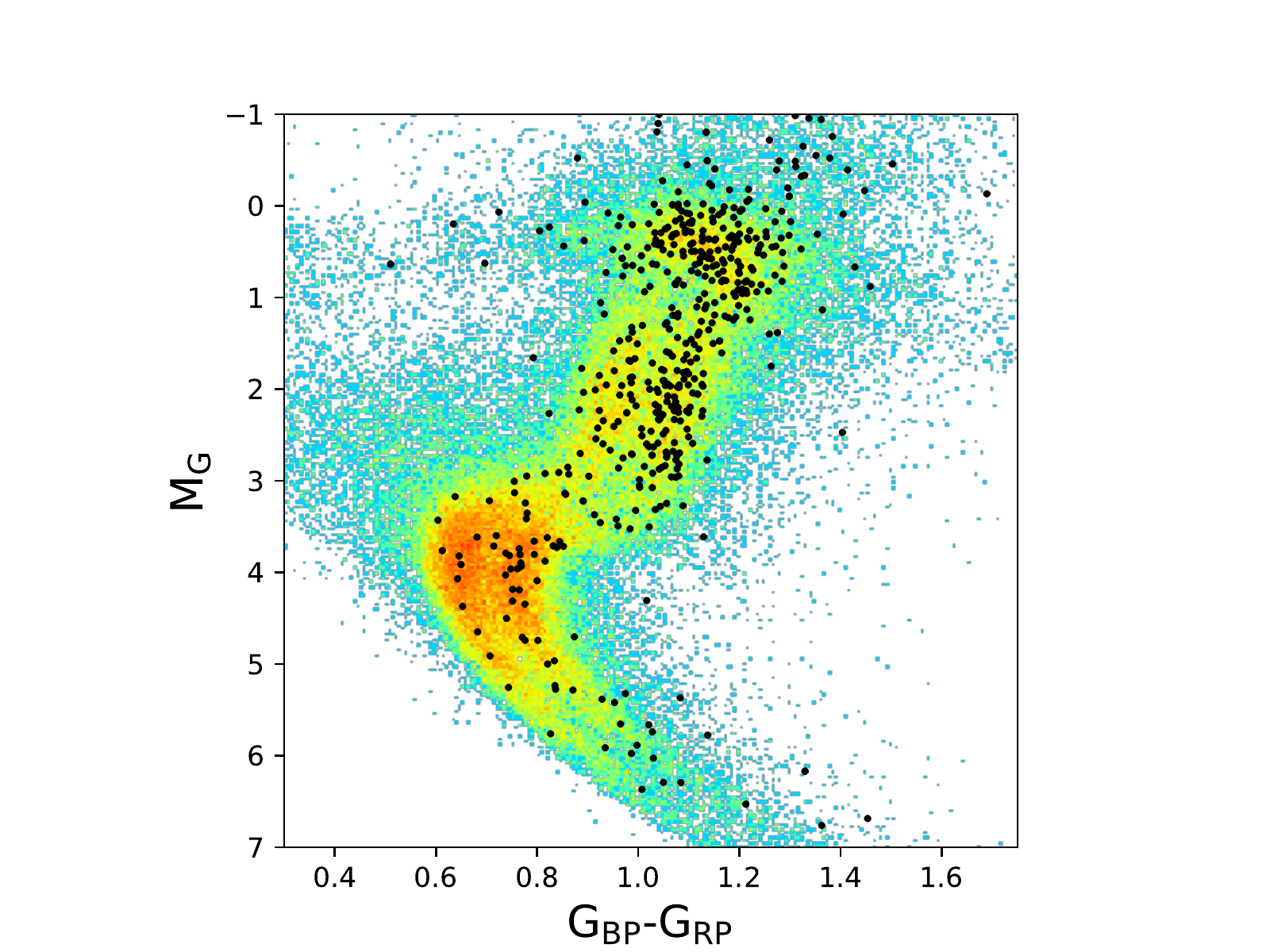}
\caption{Gaia color-magnitude diagram for the second sample described in Section \ref{sec:apogee}, 
not corrected for extinction (top) and corrected (bottom). The black dots on the bottom plot are the 950 stars found in the crossmatch
between the Gaia sample and APOGEE.}
\label{fig:dhr152742_all}
\end{figure}

\begin{figure} 
\includegraphics[width=8.5cm]{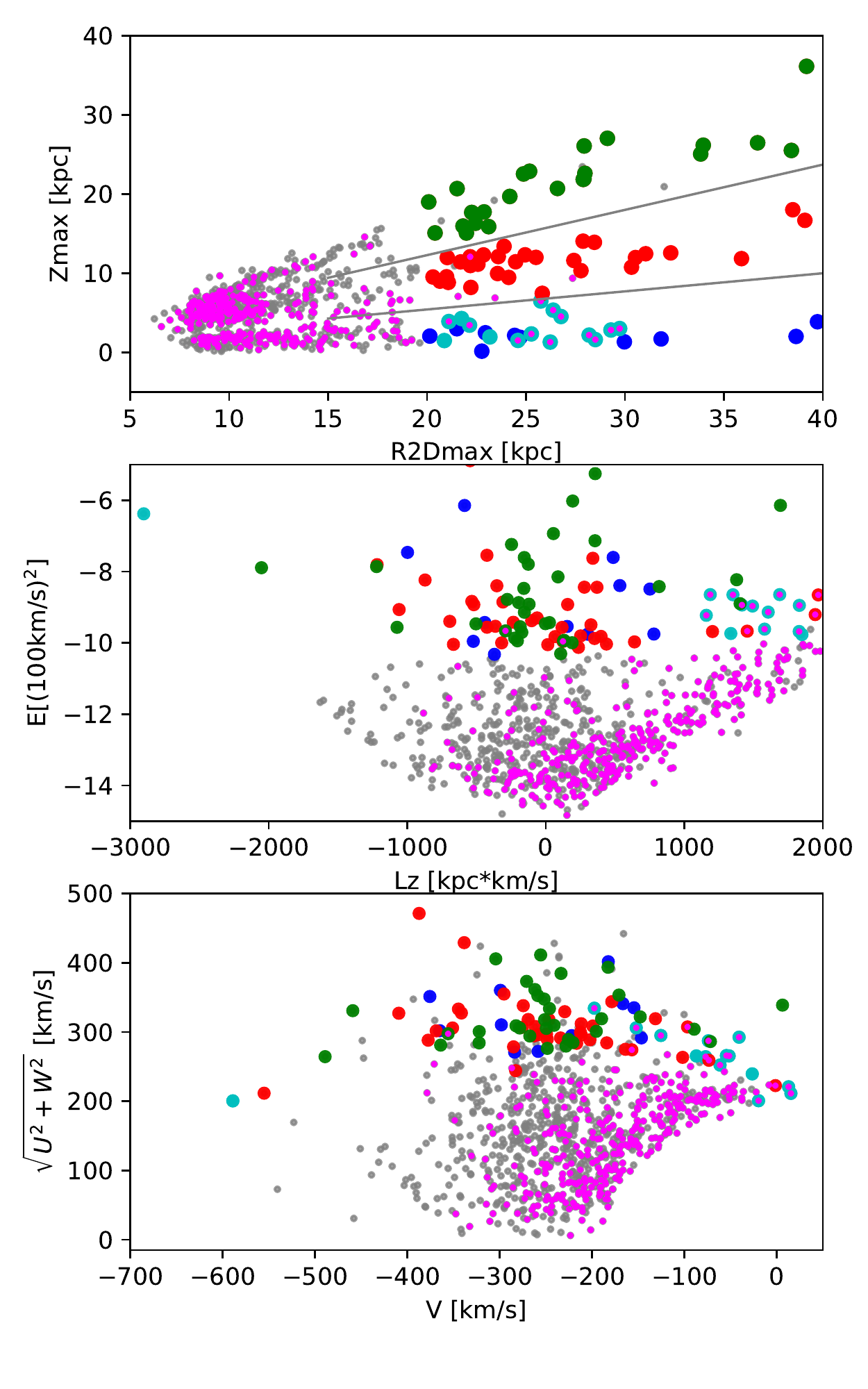}
\caption{\textit{Top:} (R2Dmax, Zmax) distribution of stars in the
APOGEE-Gaia crossmatch.  Stars belonging to the three wedges with
R2Dmax$>$20~kpc are colored in green, red, and blue.  The stars
the low altitude wedge are plotted in two different colors 
according to the pericenter: blue for stars with R2Dmin$<$2kpc,
cyan for stars with R2Dmin$>2$~kpc.  In magenta are the thick disk stars
selected to have [Fe/H]$>$-1 and [Mg/Fe]$>$0.2 in the next
figure.}\label{fig:orbacc}
\end{figure}

In order to explore further the picture presented above, there are
several limitations of our analysis so far that we can overcome. Since
our sample is limited to distances where stars have $\pi>$~1~mas,
our sample contains few giants, which are the objects with the best
S/N ratio in APOGEE.  Therefore, in order to include more giants,
we selected in the Gaia DR2 catalogue stars having $\pi>$~0.3~mas,
$\sigma_{\pi}/\pi >$~0.1 and G$<$15. The corresponding color-magnitude
diagram is shown in Fig. \ref{fig:dhr152742_all}, not corrected for
extinction and reddening (top) and corrected (bottom).  Both sequences
are well separated and densely populated and contain considerably more
giants, as expected.  Crossmatching this Gaia sample with APOGEE yields
950 stars with abundance estimates. Fig.~\ref{fig:orbacc} gives the
(R2Dmax, Zmax), (E,Lz) distributions and Toomre diagram of these 950 stars,
as calculated in the AS potential.  As expected, the new sample follows
closely the distribution of Fig.~\ref{kins}.

Based on what we learned in the previous section, we now select objects
with R2Dmax$>$20~kpc on the three `wedges' colored in blue, red and green,
assuming these are the most likely to have been accreted, with the aim to
see how they distribute in the [Fe/H]-[Mg/Fe] plane. In the lower altitude
wedge, stars are represented by two symbols according to the value of
their pericenter: below or above 2~kpc in blue or cyan respectively.

\begin{figure} 
\includegraphics[width=9.cm]{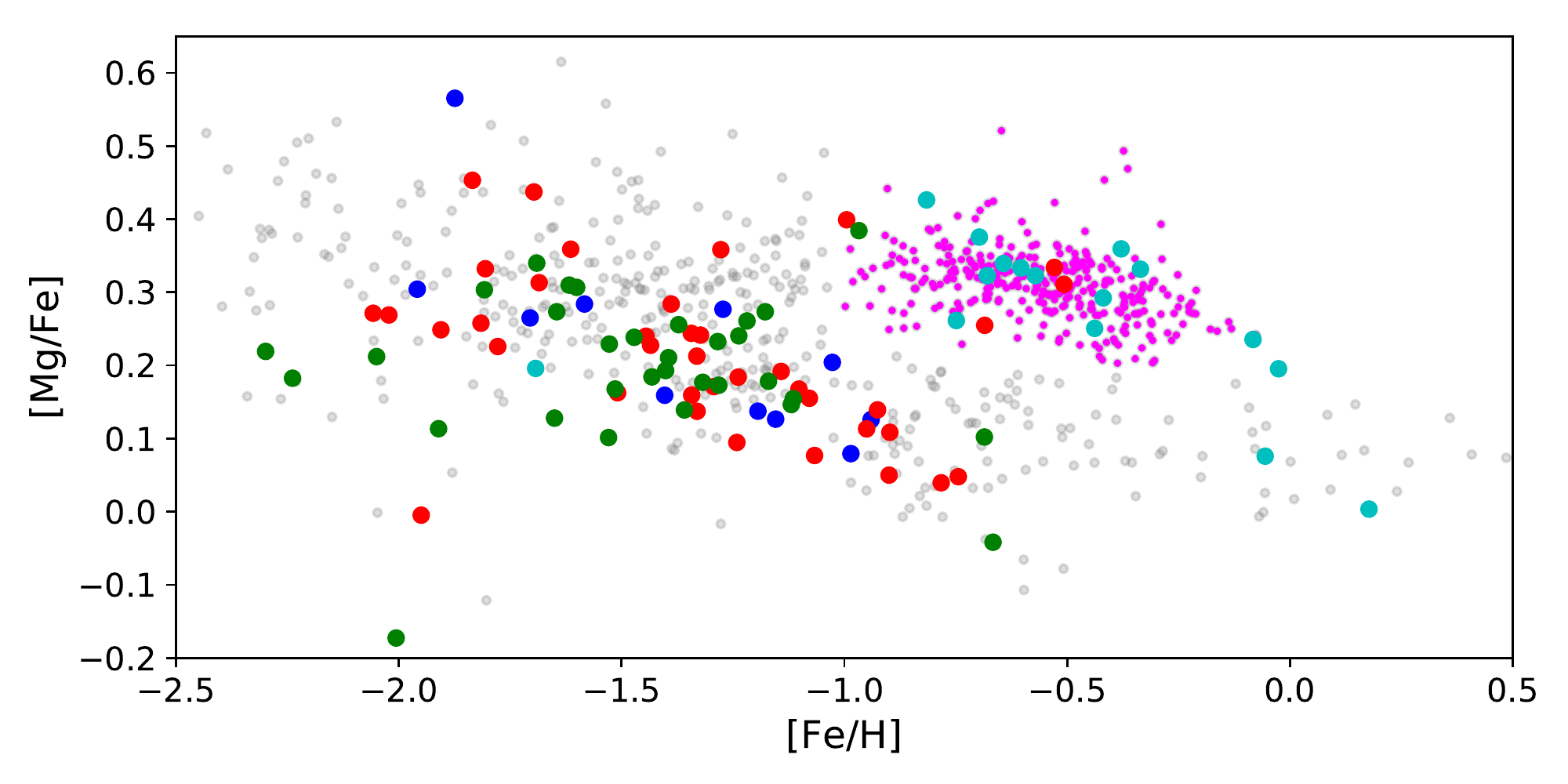}
	\caption{[Fe/H]-[Mg/Fe] distribution of stars of the APOGEE-Gaia sample. The stars selected on the basis 
	of their R2Dmax (blue, red, green) are the same as in Fig. \ref{fig:orbacc}. Thick disk stars are represented by 
	magenta points.}\label{fig:fehmg}
\end{figure}

\begin{figure}
\includegraphics[clip=true, trim = 20mm 0mm 10mm 13mm,width=1.1\linewidth]{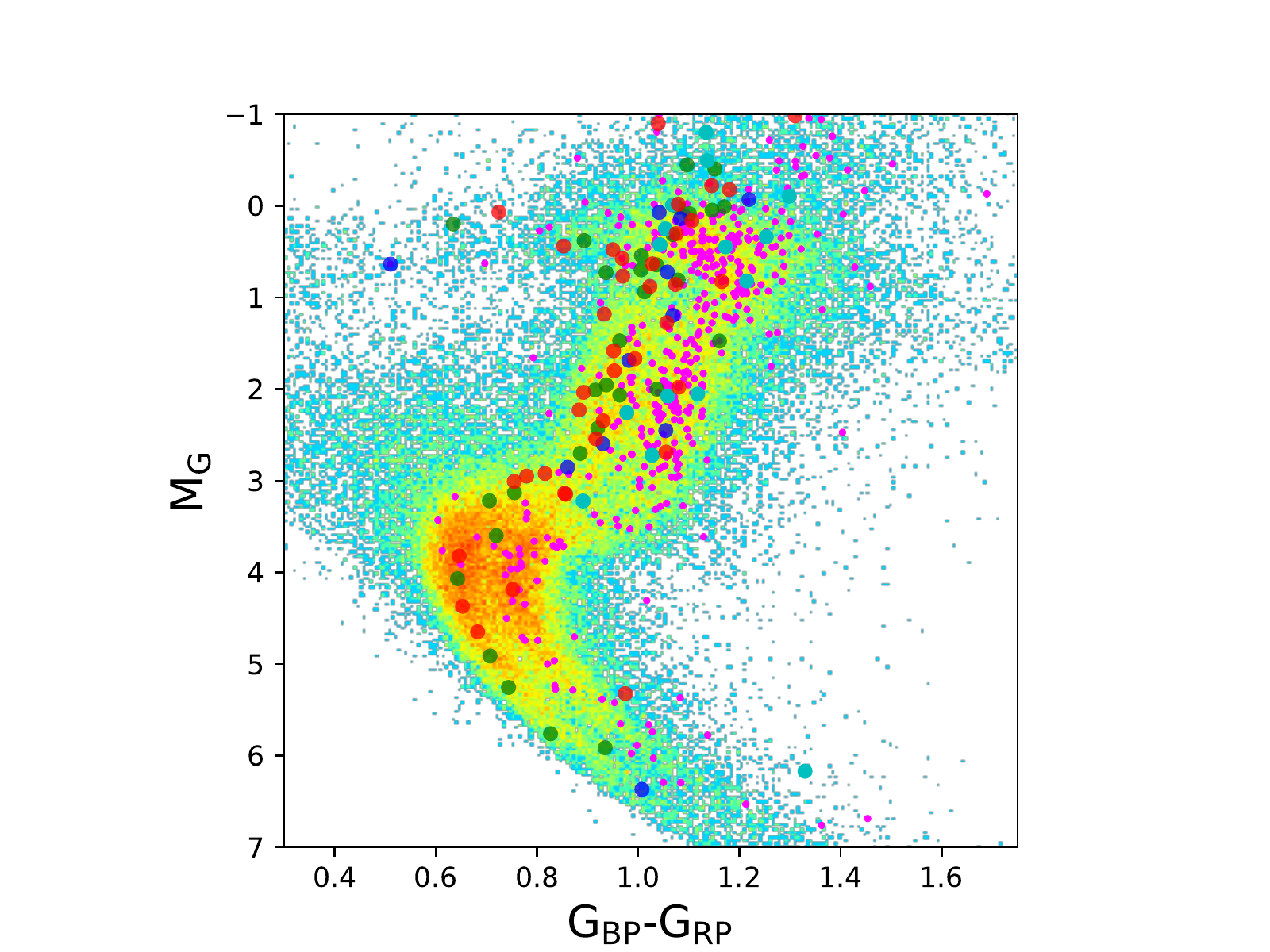}
	\caption{Gaia colour-magnitude diagram for the APOGEE stars selected as described in Sect. \ref{sec:apogee}. Symbols are the same 
	as in Fig. \ref{fig:orbacc}. Magnitudes have been corrected for extinction.
	}\label{fig:dhr152742}
\end{figure}

The [Fe/H]-[Mg/Fe] distribution of these different subsamples is given
in Fig. \ref{fig:fehmg}. It shows that stars with large apocenters
(high energy orbits) clearly comes out as forming a low star formation
efficiency sequence, typical of a relatively massive dwarf galaxy,
extending from [Fe/H]$\sim$ -0.7 to [Fe/H]$\sim$ -2.0.

Stars in the low altitude wedge are a mix of different origins: those
colored in blue also belong to the low star formation efficiency sequence,
and they differ from the others (coloured in cyan) by having pericenters
smaller than 2 kpc and which bring them near to the central regions.
The objects colored in cyan have larger pericenters, typically between
3 and 7 kpc, and are mostly thick disk stars
(Fig. \ref{fig:fehmg}). There represent the tail of this population
scattered at large galactocentric distances.

Figure \ref{fig:orbacc} shows as magenta dots stars that are chemically
defined as thick disk objects in Fig. \ref{fig:fehmg}. It confirms that
bona fide thick disk stars have a rotation which decreases to zero,
and some are counter-rotating, up to relatively high metallicities
([Fe/H]$>$-0.5).

Finally, Fig. \ref{fig:dhr152742} shows the stars with large
apocenters and those belonging to the thick disk overplotted on
the color-magnitude diagrams of the Gaia sample.  All stars have been
corrected for extinction.  The figure confirms the previous analysis by
showing that the blue sequence is dominated by stars that are on the
accreted sequence, while the red sequence is dominated by thick disk
objects, including those which have large apocenters but clearly belong
to the chemically defined thick disk population.

\section{Discussion and Conclusions}\label{sec:discussion}

Our results come in the context of a long history of halo
population studies.  The dichotomy observed in the Gaia HRD has been known
for several decades as due to the halo and thick disk. \cite{gilmore1985}
found that star counts at the pole are dominated by two populations with
main-sequence turn-offs dominated by stars at [Fe/H]-1.5, and the other
at [Fe/H]=-0.7.  This is also what was determined in the analysis
of Sloan data \citep{ivezic2008}.  More recently, \cite{jofre2011} studied 
the main sequence turn-off of old populations using SDSS data and determined 
the age of the inner Galactic halo. 
The flattened distribution found here, which reflects a strongly anisotropic velocity distribution, is
reminiscent of the findings of \cite{sommer1990} and echoes the analysis of \cite{chiba2000}.
After their results, the Galactic halo sampled at the solar 
vicinity has been seen as an inner, flattened and \textit{in situ} counterpart of a more extended, spherical, 
and accreted halo, a view that has been supported by other, more recent studies 
\citep[see][]{carollo2007,carollo2010,beers2012}, but it must be noted that already \cite{brook2003} proposed that 
eccentric stars in the sample of \cite{chiba2000} could have been the result of an accretion.
The main result of our study is that the inner halo is probably 
mainly composed of accreted material left by the last significant merger in our Galaxy.

Our results confirm that the red sequence must be dominated by
thick disk stars with metallicities between $-$0.4 and $-$1,
as found in \cite{GaiaBabusiaux2018}. The majority of RS stars
are on prograde orbits. However, an important, new result is that a
non-negligible fraction of RS stars are on retrograde orbits.  This can
be at least partially attributed, in figure \ref{kins}, to an imperfect
separation of stars of the BS and RS in the HRD -- that is, part of
the RS stars with very low or retrograde tangential velocities may
be in fact BS stars which have been incorrectly assigned to the RS.
Importantly, the analysis of the APOGEE sample clearly shows that
some stars defined chemically as thick disk objects have retrograde
orbits. It is known that \textit{in situ} stellar disks, dynamically
heated by one or several satellite accretion events, will also have
counter-rotating stars \citep[see][]{qu2011,jeanbaptiste2017}. Hence,
finding some non-rotating or counter-rotating stars among an \textit{in
situ} old disk population is not surprising. Although they have chemical
characteristics typical of the thick disk, RS stars have the kinematics
of halo stars \citep[see also][]{bonaca2017, posti2017}.  The most likely
origin of the red-color sequence is thus related to the old Galactic disk,
partially heated during some past accretion event(s).

\citet{GaiaBabusiaux2018} suggest that the blue-color sequence may
be related to the in-situ halo. We find here that this sequence is most probably 
dominated by accreted objects, as also suggested by \cite{koppelman2018}.

Low metallicity, [Fe/H]$<$-1.1,
high-$\alpha$ stars from \cite{nissen2010} and from the APOGEE sample studied here 
are indeed on the BS.
Our analysis shows that also low-metallicity, low-$\alpha$
stars are on this sequence (\S~\ref{sec:ns}). Based on the fact that
they belong to the same narrow sequence in the HRD and on the results
of \citet{hayes2018}, we conjecture that the high-$\alpha$ stars with
[Fe/H]$<$-1.1 could be causally connected to the higher metallicity,
low-$\alpha$ stars forming a unique chemical evolution sequence.
This is confirmed by the chemical characteristics of the stars with
the highest orbital energies in our APOGEE sample: these objects form a sequences
which extends from about [Fe/H]=-0.5 to at least -2.0 and from low
to high-$\alpha$ abundance. Most of these objects are on the blue sequence
in the Gaia HRD.

If these stars are on the same evolutionary
sequence, they must have been formed in an environment characterized
by a low star formation efficiency (i.e. long star formation
time-scale), much lower than that of the thick disk. For this
sequence, Supernovae-Ia must have started enriching the interstellar
medium in Fe at lower metallicities compared to what is observed
for thick disk stars of the Milky Way. The low-$\alpha$ stars in
the sample of \cite{nissen2010} would then only represent the 'tip
of the iceberg' of the much larger population blue-color sequence
stars studied by us and APOGEE \citep{hayes2018}.  The most natural
origin of the blue-color sequence is that it is dominated by stars
accreted from a satellite.

The stars that most likely were accreted dominate the distribution
of (Zmax,Rmax) at the largest apocenters, but it is difficult to determine
the fraction they represent at lower R2Dmax. Clearly, the low-$\alpha$
sequence is not only populated by objects having large apocenter. 
Interestingly, the sample at metallicities
lower than $-$2 have more limited excursions in R and Z, with typical orbits
being limited to within 15 and 10~kpc. This pecularity has also been noted 
by \cite{schuster2012}.

What may be the origin of these accreted stars?
\citet{belokurov2018} find a strong orbital anisotropy for stars
with metallicities above -1.7 and low anisotropy for stars below
this limit. They suggest that the majority of the halo stars within
30~kpc are the remnant of a massive satellite accreted during the
formation of the Galactic disk between about 8 and 11~Gyr ago. The
analysis of Gaia DR2 data may support this scenario for several
reasons: (1) BS stars seem to constitute a significant fraction of
all stars with high transverse velocities; (2) we show that they could
be on a chemical evolutionary track that is less $\alpha$-enriched
than disk stars at the same metallicity and thus are compatible
with an accreted population; (3) the kinematic properties of this
BS -- in particular, the high fraction of stars in retrograde orbits,
a fraction of which are confined within a flattened and extended
disk -- and more generally the discrete wedges 
in the $z_{\rm max}-R_{\rm 2D,max}$ plane -- are all
reminiscent of some impulsive heating of the early Galactic disk
related to some accretion event(s).
In this respect, the gap in the distribution of $z_{\rm max}$ values
may mark the transition from an early phase of significant stellar
accretion in the Galaxy to a more quiescent phase.  

Can all stars in the blue-color sequence be attributed to a unique
satellite accretion event, or is it possible that several satellites
contributed to make it? From the tightness of the BS in the HRD, it
is difficult to conceive that this sequence consists of populations
formed in several satellite galaxies, unless their chemical and age
properties at the time of their accretion were remarkably similar.
Also the kinematics of stars in this sequence may be compatible
with a unique -- and relatively massive -- merger. Fig.~C.3 in
\citet{jeanbaptiste2017}, for example, shows that among several
satellites accreted onto a Milky Way-type galaxy, stars originating
in one of them (satellite $\#3$ in that plot) are distributed in the
$E-L_z$ space in a way qualitatively similar to the BS stars we have
observed, with stars having both prograde and retrograde orbits, and
a significant plume of stars with $L_z$ centered around $L_z=0$, and
extending vertically to high energies. These simulations support the
notion that the blue-color sequence is (at least partially) the remnant
of a significant accretion event in the early history of the Milky Way.
Note that \citet{nissen2010, schuster2012} have evoked the possibility
that the low-$\alpha$ stars in their work may originate from $\omega$
Cen, and at this stage we cannot reject or confirm this suggestion for
the origin of the blue sequence.  \cite{koppelman2018} also note
that previous studies have associated this region to possible debris from
$\omega$ Cen. However, we note that $\omega$ Cen seems to have a peculiar
barium abundance \citep{majewski2012} which appears not compatible with
that of the low-$\alpha$ stars discussed here \citep[see][]{nissen2011}.

While our work represents only a first exploration of the stellar
halo in Gaia~DR2, it is clear that this mission, together with large
spectroscopic surveys, is reshaping the boundaries that we had
for decades assigned to the various stellar populations of the Milky Way.  Our
results suggest that what has been defined as the stars
of the \textit{in situ} stellar halo of the Galaxy may be in fact
fossil records of its last significant merger. Stars kinematically defined
to belong to the inner halo comprise this possible accreted population
and a sizable fraction of more metal-rich, [Fe/H]$\gtrsim$$-$1,
stars that are possibly the vestiges of the early disk of the Galaxy
after it was heated by one or more merger events. But where is the
parent population of the thick disk?  It is possible that this
progenitor, the \textit{in situ}, chemically-defined halo is lurking in the blue
sequence, but is under-represented in the volume probed by our
study? So the question remains, is the \textit{in situ} halo stellar
population expected in galaxy evolution models disguised among the
blue-color sequence or is it still beyond our reach?\\

Note: A month after this article in its first version was submitted to this Journal and posted on arXiv, 
Helmi et al. posted an article which confirms some of our results.

\acknowledgments
We would like to thank Poul Nissen and Timothy Beers for valuable comments, the referee for several suggestions 
which improved this work, and David Katz and Nicolas Leclerc for helping with the use of the Gaia Archive.
The Agence Nationale de la Recherche (ANR) is acknowledged for its financial support through the MOD4Gaia project (ANR-15-CE31-0007, P. I.: P. Di Matteo), also providing the postdoctoral grant for Sergey Khoperskov.
This work has made use of data from the European Space
Agency (ESA) mission Gaia (https://www.cosmos.esa.int/gaia),
processed by the Gaia Data Processing and Analysis Consortium
(DPAC, https://www.cosmos.esa.int/web/gaia/dpac/consortium). Funding
for the DPAC has been provided by national institutions, in particular
the institutions participating in the Gaia Multilateral Agreement.
This research has made use of the SIMBAD database,
operated at CDS, Strasbourg, France.
Funding for the Sloan Digital Sky Survey IV has been provided by the Alfred P. Sloan Foundation, the U.S. Department of Energy Office of Science, and the Participating Institutions. SDSS-IV acknowledges
support and resources from the Center for High-Performance Computing at
the University of Utah. The SDSS web site is www.sdss.org.
SDSS-IV is managed by the Astrophysical Research Consortium for the 
Participating Institutions of the SDSS Collaboration including the 
Brazilian Participation Group, the Carnegie Institution for Science, 
Carnegie Mellon University, the Chilean Participation Group, the French Participation Group, Harvard-Smithsonian Center for Astrophysics, 
Instituto de Astrof\'isica de Canarias, The Johns Hopkins University, 
Kavli Institute for the Physics and Mathematics of the Universe (IPMU) / 
University of Tokyo, Lawrence Berkeley National Laboratory, 
Leibniz Institut f\"ur Astrophysik Potsdam (AIP),  
Max-Planck-Institut f\"ur Astronomie (MPIA Heidelberg), 
Max-Planck-Institut f\"ur Astrophysik (MPA Garching), 
Max-Planck-Institut f\"ur Extraterrestrische Physik (MPE), 
National Astronomical Observatories of China, New Mexico State University, 
New York University, University of Notre Dame, 
Observat\'ario Nacional / MCTI, The Ohio State University, 
Pennsylvania State University, Shanghai Astronomical Observatory, 
United Kingdom Participation Group,
Universidad Nacional Aut\'onoma de M\'exico, University of Arizona, 
University of Colorado Boulder, University of Oxford, University of Portsmouth, 
University of Utah, University of Virginia, University of Washington, University of Wisconsin, 
Vanderbilt University, and Yale University.

\end{document}